\documentclass{IEEEojcsys}

\usepackage[colorlinks,urlcolor=blue,linkcolor=blue,citecolor=blue]{hyperref}
\usepackage{color,array}

\usepackage{graphicx}

\usepackage[utf8]{inputenc} 
\usepackage[T1]{fontenc}    
\usepackage{hyperref}       
\usepackage{amsmath}
\usepackage{url}            
\usepackage{amsfonts}       
\usepackage{lipsum}
\usepackage{tikz}
\usetikzlibrary{calc,positioning,decorations.pathreplacing}
\usepackage{algorithm}
\usepackage{algorithmic}
\usepackage{amssymb}

\newtheorem{Theorem}{Theorem}

\newtheorem{Lemma}{Lemma}
\newtheorem{Remark}{Remark}
\newtheorem{Corollary}{Corollary}

\newenvironment{proof}[1][Proof]{\par\noindent\textbf{#1.} }{\hfill\par}
\jvol{00}
\jnum{XX}
\paper{1234567}
\pubyear{2021}
\receiveddate{XX September 2021}
\accepteddate{XX October 2021}
\publisheddate{XX November 2021}
\currentdate{XX November 2021}
\doiinfo{OJCSYS.2021.Doi Number}

\begin{document}

\sptitle{Article Category}

\title{Finite-Sample Analysis of Model-free Policy Gradient Methods for LQR with Stochastic Data} 

\editor{This paper was recommended by Associate Editor F. A. Author.}

\author{Bowen Song\affilmark{1}}

\author{Andrea Iannelli\affilmark{1}  (Member, IEEE)}

\affil{Institute for Systems Theory and Automatic Control, University of Stuttgart, Germany} 

\corresp{CORRESPONDING AUTHOR: Bowen Song (e-mail: \href{mailto:bowen.song@ist.uni-stuttgart.de}{bowen.song@ist.uni-stuttgart.de})}
\authornote{Bowen Song acknowledges the support of the International Max Planck Research School for Intelligent Systems (IMPRS-IS). }

\markboth{Finite-Sample Analysis of Model-free Policy Gradient Methods for LQR with Stochastic Data}{Bowen Song {\itshape ET AL}.}

\begin{abstract}
Policy gradient (PG) methods are the backbone of many reinforcement learning algorithms due to their good performance in policy optimization problems. As a gradient-based approach, PG methods typically rely on knowledge of the system dynamics. If this is not available, trajectory data can be utilized to approximate first-order information. When the data are noisy, gradient estimates become inaccurate and a study that investigates uncertainty estimation and the analysis of its propagation through the algorithm is currently missing. To address this, our work focuses on the Linear Quadratic Regulator (LQR) problem for systems subject to additive stochastic noise. After briefly summarizing the state of the art for cases with a known model, we focus on scenarios where the system dynamics are unknown, and approximate gradient information is obtained using zeroth-order optimization techniques. We analyze the theoretical properties by computing the error in the estimated gradient and examining how this error affects the convergence of PG algorithms. \textcolor{black}{Additionally, we provide global convergence guarantees for various versions of PG methods, including those employing adaptive step sizes and variance reduction techniques.} This study contributes to characterizing the robustness of model-free PG methods, aiming to identify their limitations in the presence of stochastic noise and proposing improvements to enhance their applicability.
\end{abstract}


\maketitle

\section{INTRODUCTION}

Reinforcement learning \cite{Annaswamy_23_ARCRAS,bertsekas2019reinforcement,Sutton1998} has demonstrated a significant influence across a wide range of applications. A key concept within reinforcement learning is policy optimization, whereby the policy is parameterized and directly optimized based on a predefined performance metric \cite{annurev:/content/journals/10.1146/annurev-control-042920-020021}. Several successful policy optimization methods have been developed, including policy gradient \cite{10.5555/3009657.3009806}, actor-critic \cite{NIPS1999_6449f44a}, and proximal policy optimization \cite{schulman2017proximalpolicyoptimizationalgorithms}. This work focuses on policy gradient (PG) methods, based on the idea of minimizing a cost function over the parameterized policy through a gradient descent-type update. Studying the convergence properties of PG methods, particularly their global convergence to the optimal policy, is an active area of research \cite{cen2023globalconvergencepolicygradient,pmlr-v80-fazel18a,9254115,doi:10.1137/20M1347942,doi:10.1137/19M1288012,10005813,9992612,doi:10.1137/23M1560781}. For instance, PG methods have been applied in \cite{doi:10.1137/20M1347942} to solve $\mathcal{H}_2$ cost subject to $\mathcal{H}_\infty$ constraints, and in \cite{doi:10.1137/19M1288012} to Markov decision processes.

The application of reinforcement learning to the linear quadratic regulator (LQR) problem has been extensively explored due to its value as an analytically tractable benchmark for studying reinforcement learning with continuous state and action spaces \cite{pmlr-v54-abeille17b,articlesimulation,pmlr-v80-fazel18a,JMLR:v24:21-0842,8558117,10384256,annurev:/content/journals/10.1146/annurev-control-053018-023825,pmlr-v80-tu18a}. As a consequence, PG methods have gained attention in recent studies on the LQR problem, increasing significantly after \cite{pmlr-v80-fazel18a} demonstrated global convergence of PG applied to the deterministic LQR problem. The work \cite{10005813} explores a primal-dual policy gradient method for the constrained LQR problem, while \cite{9992612} studies the performance limitations of PG applied to LQR from a control-theoretic perspective.

Solving the LQR problem requires the availability of a system model, as assumed in \cite{doi:10.1137/20M1347942,doi:10.1137/19M1288012,10005813,9992612}. However, in real-world applications a complete system description is often unavailable, making it necessary to combine policy optimization with data-driven techniques. To address this, various approaches have been developed. Methods like \cite{pmlr-v19-abbasi-yadkori11a,pmlr-v119-cassel20a,pmlr-v119-simchowitz20a} combine system identification with model-based LQR design and use regret as a performance metric. The same two-step approach can be used with gradient-based schemes: \cite{https://doi.org/10.1002/rnc.7475} integrates recursive least squares with policy iteration, while \cite{10383604} combines recursive least squares with policy gradient methods. Both \cite{https://doi.org/10.1002/rnc.7475,10383604} assume noise-free data, whereas \cite{SongIannelli+2025+398+412,zhao2025policygradientadaptivecontrol} extend these frameworks to bounded-noise settings. These are termed indirect data-driven methods, since data is first used to estimate a model, which is then integrated into a model-based certainty equivalence design.

Direct data-driven methods instead use data directly to determine the policy, bypassing the intermediate model estimation process \cite{11030798}. In \cite{pmlr-v80-fazel18a}, model-free PG methods use zeroth-order optimization to estimate gradients for the LQR cost from finite, noise-free trajectories. The work \cite{carnevale2025datadrivenlqrfinitetimeexperiments} proposes a novel gradient estimation method for noise-free systems. \cite{9448427} uses PG to solve the continuous-time state feedback LQR problem, and \cite{10091214} applies PG to a continuous-time output feedback system using zeroth-order gradient estimation with variance reduction. \cite{10669082} focuses on stabilizing the unknown system via policy gradient. In all these direct data-driven works \cite{pmlr-v80-fazel18a,carnevale2025datadrivenlqrfinitetimeexperiments,9448427,10091214,10669082}, only the noise-free case is studied. However, it is essential to analyze what happens when gradients and other objects in the policy update are estimated from noisy data. In \cite{9254115}, model-free PG is applied to linear systems with multiplicative noise, addressing uncertainty in the system matrix. The works \cite{9254115,10091214,10669082} all build on \cite{pmlr-v80-fazel18a}, leveraging the inherent robustness of PG methods: informally, when the gradient estimation error is sufficiently small, one can still guarantee a decrease in the cost function. However, in \cite{pmlr-v80-fazel18a,9254115,10091214,10669082}, stochastic uncertainty is not considered, and the analysis assumes strictly bounded trajectory data. Another line of research tackles the model-free LQR problem differently: \cite{pmlr-v89-malik19a} derives convergence properties by carefully selecting step sizes under gradient uncertainty, and follow-up works \cite{li2025robustnessderivativefreemethodslinear,moghaddam2025samplecomplexitylinearquadratic} refine this with alternative gradient estimation techniques, yielding improved sample efficiency and robustness guarantees. These results \cite{pmlr-v89-malik19a,li2025robustnessderivativefreemethodslinear,moghaddam2025samplecomplexitylinearquadratic} assume stochastic but bounded noise. Fewer studies consider stochastic unbounded noise: \cite{doi:10.1137/20M1382386} applies PG to finite-horizon LQR \textcolor{black}{with a time-varying optimal policy}, while \cite{doi:10.1137/23M1554771,NEURIPS2019_9713faa2} propose model-free first-order methods that estimate individual gradient components via least squares, differing from the zeroth-order approaches above. To date, there is no comprehensive analysis that jointly addresses the uncertainty from zeroth-order estimation and its propagation through the algorithm while providing probabilistic convergence guarantees and sample complexity bounds for the infinite-horizon LQR problem.

In this work, we analyze the sample complexity and robustness to noise of PG methods for the model-free LQR with additive stochastic noise. Due to the unbounded support of the noise, we consider an average infinite-horizon cost. We first characterize the properties of this cost function and analyze model-based versions of a few PG algorithms. \textcolor{black}{For the policy gradient descent (PGD) and natural policy gradient (NPG) methods, we propose an adaptive stepsize scheme with a provably no slower convergence rate than the fixed step size methods in \cite{pmlr-v80-fazel18a,9254115,10091214}, and demonstrate numerically that it achieves faster convergence in practice.} Then, for both model-free PGD and NPG, we employ zeroth-order optimization to approximate the gradient and covariance from noisy data. To handle the challenge of unbounded gradient estimates due to stochastic noise, we introduce an event-based analysis, extending \cite{doi:10.1137/20M1382386} while relaxing the reliance on gradient upper bounds. We provide convergence guarantees for these model-free PG methods with a sample complexity accounting for the stochastic nature of the estimation error. \textcolor{black}{Furthermore, we introduce and analyze a variance reduction technique within the zeroth-order optimization framework, proving that it achieves a smaller sample complexity than without variance reduction, with practical improvements further confirmed by numerical experiments. Although variance reduction for model-free LQR was already presented in \cite{10091214}, our contribution here is new, as we deal with stochastic data and provide improved sample complexity guarantees.} Finally, we compare the guarantees of model-free PG methods with and without noisy data, demonstrating the importance of considering noise and offering guidance on parameter tuning in applications. Our analysis shows that model-free PG with stochastic data enjoys qualitatively similar convergence guarantees to the noise-free scenario, but points out limitations in terms of the additional samples required for a given accuracy, reduced step size ranges, and convergence rate.

\subsection*{Notations}
We denote by $A\succeq 0$ and $A\succ0$ a positive semidefinite and positive definite matrix $A$, respectively. The symbol $\lambda_1(A)$ denotes the smallest eigenvalue of the matrix $A$. For matrices, $\lVert \cdot\rVert_F$ and $\lVert \cdot\rVert$ denote the Frobenius norm and induced $2$-norm, respectively. For vectors, $\lVert \cdot\rVert$ denotes the Euclidean norm. $I$ represents the identity matrix. $\mathbb{Z}_+$ is the set of non-negative integers. The operation $vec(A)=[a_1^\top,a_2^\top,...,a_n^\top]^\top$ stacks the columns of matrix $A$ into a vector. $\mathcal{N}(0,\Sigma)$ denotes a Gaussian distribution with $0$ mean and covariance $\Sigma \succ 0$. \textcolor{black}{We write $a = \mathcal{O}(b)$ for positive scalars $a, b$ if there exists a constant $c > 0$ such that $a \leq c \cdot b$.}

\section{Preliminaries}\label{Section2}
\label{sec:Preliminaries}
In this work, we consider the following averaged infinite horizon optimal control problem,  where the plant is subject to additive stochastic noise:
\vspace{-3pt}
\begin{subequations}\label{2}
\begin{align}
        \min_{u_t} \lim_{T\rightarrow +\infty}&\frac{1}{T}\mathop{\mathbb{E}}_{x_0, w_t} \sum_{t=0}^{T-1}\left( x_t^\top Q x_t +u_t ^\top R u_t\right),\label{Cost}\\ 
        \mathrm{s.\,t.} ~ &x_{t+1}=Ax_t+Bu_t+w_t, \label{LTI}\\&w_t \sim \mathcal{N}(0,\Sigma_w),\textcolor{black}{x_0 \sim \mathcal{N}(0,\Sigma_0)},\label{i}
\end{align}
\end{subequations}
where $x_t\in \mathbb{R}^{n_x}$ is the system state and $u_t\in \mathbb{R}^{n_u}$ is the system input; $A\in \mathbb{R}^{n_x \times n_x}$, $B\in \mathbb{R}^{n_x \times n_u}$, $(A,B)$ is unknown but stabilizable, which is a standard assumption in data-driven control \cite{pmlr-v99-tu19a,9691800}; $Q,R\succ 0$ are the weighting matrix. We define the state covariance at $t=0$ as $\Sigma_0:=\mathop{\mathbb{E}}\limits_{x_0}\left[x_0x_0^\top\right]$. 

\textcolor{black}{Motivated by the fact that the optimal LQR controller for problem \eqref{2} is a static linear state feedback law \cite{lewis2012optimal}, we parameterize the input as $u_t=Kx_t$ with gain $K$}. The optimal cost only depends on $K$ and is denoted by $C$:
\begin{equation}\label{CostFunction0}
    C(K):=\lim_{T\rightarrow +\infty}\frac{1}{T}\mathop{\mathbb{E}}_{x_0,w_t} \sum_{t=0}^{T-1} x_t^\top {\left(Q+K^\top R K\right)}x_t.
\end{equation}
with $Q_K:=Q+K^\top R K$.
\begin{Remark}[Weighting matrices $Q,R$]\label{RemarkQR}
Prior knowledge of the weighting matrices $Q$ and $R$ is not strictly required for implementing the model-free policy gradient algorithms investigated here. While the analyses presented here assume their knowledge, it suffices to have access to the empirical cost, which can be estimated from the closed-loop trajectories of system \eqref{LTI} used to estimate gradient information. 
\end{Remark}
\subsection{Properties of $C$}\label{section2.2}
In this section, we introduce basic properties of problem \eqref{2} that are particularly relevant when policy gradient methods are used to solve it. While these results are mostly well-known, they form the foundation for the main results of this work in Sections \ref{sec:MBGD} and \ref{sec:MFGD}.  The state response $x_t$ for $t\geq 1$ associated with any gain $K$ is given by:
\textcolor{black}{
\vspace{-3pt}
\begin{equation} \label{stateresponce}
    x_t=(A_K)^tx_0+\sum_{k=0}^{t-1} (A_K)^{t-k-1}w_{k}, \quad \forall t\geq 1.
\end{equation}}
We define the set $\mathcal{S}$ as the set of matrices $K\in \mathbb{R}^{n_x \times n_u}$ that stabilizes the system $(A,B)$, meaning that the matrix $A+BK$ is Schur stable:
\vspace{-3pt}
\begin{equation*}
    \mathcal{S}:=\left\{K \in \mathbb{R}^{\textcolor{blue}{n_u \times n_x}} | \rho (A_K ) <1,~\mathrm{with}~A_K:=A+BK \right\}.
\end{equation*}

Now, we introduce some important definitions associated with $K \in \mathcal{S}$. The covariance matrix at time $t$ is defined as $\Sigma_t:=\mathop{\mathbb{E}}\limits_{x_0,w_t}[x_t x_t^\top]$. From this definition and the state response given in \eqref{stateresponce}, we obtain:
\begin{equation}\label{covariance}
           \Sigma_{t+1}=A_K\Sigma_{t}A_K^\top+\Sigma_w,~t\in \mathbb{Z}_{+},\\
\end{equation}
\textcolor{black}{with~$\Sigma_0$ introduced in \eqref{i}}. The average covariance matrix associated with $K\in \mathcal{S}$ is defined as:
\begin{equation}\label{definedAverage}
    \Sigma_K:=\lim_{T\rightarrow +\infty}\frac{1}{T}\sum_{t=0}^{T-1} \Sigma_t=\sum_{t=0}^{\infty}A_K^t \Sigma_w A_K^{t~\top}.
\end{equation}
The cost function $C(K)$ associated with $K\in \mathcal{S}$ can be equivalently expressed as:
\begin{equation}\label{CostFunction}
C(K)=\mathrm{Tr}\left(P_K\Sigma_w\right)=\mathrm{Tr}\left(Q_K\Sigma_K\right),
\end{equation}
where $P_K$ is defined as the solution of the following Lyapunov equation:
\begin{equation*}
\begin{split}
    P_K=Q_K+A_K^\top P_K A_K=\sum_{t=0}^{\infty}A_K^{t~\top} Q_K A_K^{t}.
\end{split}
\end{equation*}
From the expression of the cost function \eqref{CostFunction} and well-known results \cite{lewis2012optimal}, the optimal $K^*$, which minimizes the cost function $C$, is given by:
\begin{equation*}
    \begin{split}
K^*&=-(R+B^\top P_{K^*}B)^{-1}B^\top P_{K^*}A,  \\
P_{K^*}&=Q+A^\top P_{K^*}A-\\
&~~~~~~~~A^\top P_{K^*}B(R+B^\top P_{K^*}B)^{-1}B^\top P_{K^*}A. 
    \end{split}
\end{equation*}
The average covariance matrix associated with the optimal $K^*$ is denoted as $\Sigma_{K^*}$. Using \eqref{CostFunction}, the gradient of $C(K)$ with $K\in \mathcal{S}$ can be expressed as follows \cite{9992612}:
    \begin{equation*}\label{Gradient}    
    \nabla C(K)=2E_K \Sigma_K,
\end{equation*}
where $E_K:=\left(R+B^\top P_K B\right) K+B^\top P_K A$ and $\Sigma_k$ defined in \eqref{definedAverage} satisfies:
\textcolor{blue}{\begin{equation}\label{cSigmaK}
    \Sigma_K=\Sigma_w+A_K\Sigma_KA_K^\top
\end{equation}}
It is known that the cost function $C$ defined in \eqref{CostFunction0} is generally non-convex \cite{pmlr-v80-fazel18a}. In the context of non-convex optimization, the convergence to the optimal solution of gradient descent cannot usually be guaranteed. However, the cost function $C(K)$ satisfies a special property known as \emph{gradient domination} \textcolor{black}{\cite[Lemma 3]{pmlr-v80-fazel18a}} and \emph{almost smoothness} \textcolor{black}{\cite[Lemma 6]{pmlr-v80-fazel18a}}. \textcolor{black}{These properties form the basis for establishing convergence guarantees of policy gradient methods. For brevity, we do not restate the corresponding definitions here. For later use, however, we recall the boundedness of $\lVert K \rVert$ and $\nabla C(K)$, and local Lipschitz continuity property of the $C,\Sigma_k,\nabla C$.}


\begin{Lemma}[Boundedness of $\lVert\nabla C(K)\rVert, \lVert K\rVert $] \label{bound}
Given any $K\in  \mathcal{S}$, we have
\begin{equation*}
    \textcolor{black}{\begin{split}
        \lVert \nabla C(K) \rVert_F\leq b_{\nabla}(C(K)),\quad \lVert K \rVert \leq b_K(C(K)),
    \end{split}}
\end{equation*}
    where the expressions for $b_\nabla$ and $b_K$ are given in \eqref{boundedgradienteq} in Appendix \ref{DetailedExpression}.
\end{Lemma}
\begin{Lemma}[Lipschitz continuity of $\Sigma_K,C,\nabla C$]\label{Lipschitz}
    Suppose $K',K \in \mathcal{S}$ are such that:
    \begin{equation*}\label{SigmaK1}
    \lVert K-K' \rVert \leq \frac{\lambda_1(\Sigma_w)\lambda_1(Q)}{4C(K)\lVert B \rVert(\lVert A+BK\rVert+1)}=: h(C(K)),
\end{equation*}
it holds that:
\begin{equation*}\label{SigmaK2}
    \lVert \Sigma_K-\Sigma_{K'} \rVert \leq{h_{\Sigma}(C(K))}\lVert K-K' \rVert,
\end{equation*}
with ${h_{\Sigma}(C(K))}:={4 \left(\frac{C(K)}{\lambda_1(Q)} \right)^2 \frac{\lVert B \rVert(\lVert A+BK\rVert+1)}{\lambda_1(\Sigma_w)}}$.
If $ \lVert K-K' \rVert \leq\min \{h\left(C(K)\right), \lVert K^* \rVert\},$
    it holds that:
    \begin{equation*}
    \begin{split}
        \lVert C(K)-C(K')\rVert \leq h_{C}(C(K))\lVert K-K' \rVert,\\
        \lVert \nabla C(K)-\nabla C(K') \rVert \leq h_{\nabla}(C(K)) \lVert K-K' \rVert,
    \end{split}        
\end{equation*}
where $h_{C},h_{\nabla}$ are defined in \eqref{Pertubation1} and \eqref{ErrorGradient} in Appendix \ref{DetailedExpression}, respectively.
\end{Lemma} 
\textcolor{black}{The proofs of Lemmas \ref{bound} and \ref{Lipschitz} are straightforward extension of \cite[Lemma 16, Lemma 24 and Lemma 25]{pmlr-v80-fazel18a} by adapting the expressions of $\Sigma_K$, $C$, and $\nabla C$. }

\section{Convergence of Model-based Policy Gradient}\label{sec:MBGD}
In this section, we show the global convergence properties of model-based policy gradient method (PGM), that is, algorithms that operate with perfect knowledge of the system matrices $(A, B)$. We consider two variants of PGMs: 
\begin{itemize}
    \item Policy gradient descent:  $K_{i+1}=K_i-\eta_i \nabla C(K_i)$;
    \item Natural policy gradient: $K_{i+1}=K_i-\eta_i \nabla C(K_i) \Sigma_{K_i}^{-1}$,
\end{itemize}
where $i\in \mathbb{Z}_+$ is the iteration index and $\eta_i$ is the step size.
\subsection{Policy Gradient Descent (PGD)}
In this subsection, we present the theorem guaranteeing the convergence of the model-based policy gradient descent method, based on the properties introduced in Sections \ref{Section2}.
\begin{Theorem}[PGD with Adaptive Step Size]\label{PGTheorem}
    Suppose the initial $K_0 \in \mathcal{S}$, and consider the policy gradient descent iteration
     \begin{equation}\label{PGD1}
    K_{i+1}=K_i-\eta_i \nabla C(K_i), \quad \forall i\in \mathbb{Z}_+,
\end{equation}
where the step size satisfies $\eta_i \leq h_{\mathrm{PGD}}(C(K_i))$, and $h_{\mathrm{PGD}}(C(K_i))$ is defined in \eqref{pG2} in Appendix \ref{DetailedExpression}. Then, the following relationship holds: 
\begin{equation}\label{GD}
    C(K_{i+1})-C(K^*) \leq \gamma_i^p (C(K_{i})-C(K^*)),
\end{equation}
with $\gamma^p_i:=1-\frac{2\eta_i \lambda_1{(R)} \lambda_1^2(\Sigma_w)}{\lVert \Sigma_{K^*}\rVert}$. For $\eta_i=h_{\mathrm{PGD}}(C(K_i)), i\in \mathbb{Z}_+$ and given any accuracy gap $\epsilon >0$, if the number of iterations $N$ satisfies:
\begin{equation*}
    N \geq  \frac{\lVert \Sigma_{K^*}\rVert}{2{\eta_0} \lambda_1(R)\lambda^2_1(\Sigma_w)}\log \frac{C(K_{0})-C(K^*)}{\epsilon},
\end{equation*}
then $C(K_N)-C(K^*) \leq \epsilon.$
\end{Theorem}
\textcolor{black}{The proof of Theorem \ref{PGTheorem} is provided in Appendix \ref{Proof1}}. According to Theorem \ref{PGTheorem}, the step size is adaptive according to the cost $C(K_i)$ at the current iterate $K_i$. 
\textcolor{black}{
\begin{Remark}[Advantage of Adaptive Step Size] \label{RemarkGD}
We discuss here the advantage of using the adaptive step size proposed in Theorem~\ref{PGTheorem}. The work in \cite{pmlr-v80-fazel18a} proposes a fixed step size $\eta = h_\mathrm{PGD}(C(K_0))$, computed once from the initial cost $C(K_0)$, which guarantees the monotonic decrease $C(K_{i+1})\leq C(K_i)$ for all iterations $i$. By contrast, the adaptive step size $\eta_i = h_\mathrm{PGD}(C(K_i))$ is recomputed at every iteration and, since $C(K_i)$ is non-increasing, it grows as the iterations proceed: $\eta=\eta_0\leq \eta_i$ for all $i \in \mathbb{Z}_+$. This guarantees that the adaptive step size yields a per-iteration contraction factor no worse than the fixed step size, as shown by the following inequality obtained from \eqref{GD} for all $i\in \mathbb{Z}_+$:
 \begin{equation*}
     \begin{split}
        0<1-\frac{2\eta_i \lambda_1{(R)} \lambda_1^2(\Sigma_w)}{\lVert \Sigma_{K^*}\rVert} \leq 1-\frac{2\eta_0 \lambda_1{(R)} \lambda_1^2(\Sigma_w)}{\lVert \Sigma_{K^*}\rVert} <1.
     \end{split}
 \end{equation*}
\end{Remark}}

\subsection{Natural Policy Gradient (NPG)}
The natural policy gradient method adjusts the standard policy gradient by considering the geometry of the parameter space through the Fisher information matrix \cite{NIPS2001_4b86abe4}. 
The following results establish the convergence properties of NPG.
\begin{Theorem}[NPG with Adaptive Step Size]\label{NPGTheorem}
     Suppose the initial $K_0 \in \mathcal{S}$, and consider the natural policy gradient iteration
     \begin{equation*}
    K_{i+1}=K_i-\eta_i \nabla C(K_i) \Sigma_{K_i}^{-1}, \quad \forall i\in \mathbb{Z}_+,
\end{equation*}
where the step size satisfies $ \eta_i \leq \frac{1}{2\lVert R \rVert+\frac{2\lVert B \rVert^2 C(K_i)}{\lambda_1(\Sigma_w)}}$. Then the following relationship holds: 
\begin{equation}\label{PGConvergence}
    C(K_{i+1})-C(K^*) \leq \gamma_i^N (C(K_{i})-C(K^*)),
\end{equation}
with $\gamma_i^N:=1-\frac{2\eta_i \lambda_1{(R)} \lambda_1(\Sigma_w)}{\lVert \Sigma_{K^*}\rVert}$. For $\eta_i = \frac{1}{2\lVert R \rVert+\frac{2\lVert B \rVert^2 C(K_i)}{\lambda_1(\Sigma_w)}}$, $\forall i\in \mathbb{Z}_+,$ and given any accuracy gap $\epsilon >0$, if the number of iterations $N$ satisfies:
\begin{equation*}
    N \geq \frac{\lVert \Sigma_{K^*}\rVert}{2\lambda_1(\Sigma_w)\lambda_1(R)\eta_0}\log \frac{C(K_{0})-C(K^*)}{\epsilon},
\end{equation*}
then $C(K_N)-C(K^*) \leq \epsilon$.
\end{Theorem}
\textcolor{blue}{The proof of Theorem~\ref{NPGTheorem} is given in Appendix \ref{Proof2}.} \textcolor{black}{Similar to the observation in Remark~\ref{RemarkGD}, the adaptive step size for the NPG method yields a per-iteration contraction factor no worse than the fixed step size, since the admissible upper bound grows as the cost $C(K_i)$ decreases. This practical advantage will be further illustrated numerically in Figures \ref{fig:fig3} and \ref{fig:fig4} in Section \ref{sec:Simulation}.}
\subsection{Lack of robustness of model-based PG methods}\label{simu24}
The guarantees on PG methods reviewed in the previous sections rely on the availability of an exact system model. When this is not available, gradients and covariances are estimated from data trajectories, and for plants such as \eqref{LTI}, they will be subject to stochastic errors. \textcolor{black}{We exemplify with a practical example the performance of a PGD method when noisy gradients are used. We consider the control of the longitudinal dynamics of a Boeing 747 aircraft. The linearized dynamics are given by \cite{doi:10.1137/23M1554771}:
\begin{equation}\label{Boeing747}
\begin{split}&x_{t+1}=Ax_t+Bu_t+w_t,
\\
A = &
\begin{bmatrix}
1 & -1.13 & -0.65 & -0.807 & 1.59 \\
0 & 0.77 & 0.32 & -0.98 & -2.97 \\
0 & 0.12 & 0.02 & 0 & -0.36 \\
0 & 0.01 & 0.01 & -0.03 & -0.04 \\
0 & 0.14 & -0.09 & 0.29 & 0.76
\end{bmatrix},\\B =&
\begin{bmatrix}
89.20 & -50.17 & 1.13 & -19.35 \\
5.22 & 6.36 & 0.23 & -0.32 \\
-9.47 & 5.93 & -0.12 & 0.99 \\
-0.32 & 0.32 & -0.01 & -0.01 \\
-4.53 & 3.21 & -0.14 & 0.09
\end{bmatrix}.
\end{split}
\end{equation}
The process noise is sampled as $w_t \sim\mathcal{N}(0,10^{-1}I_5)$. The weight matrices $Q$ and $R$ are set to identity matrices. The initial control gain $K_0$ is chosen as the optimal solution to the LQR problem with cost matrices $(A,B,40Q,R)$. To qualitatively capture the effect of noisy data on the gradient errors,} we define the gradient estimate as follows:
\begin{equation}\label{arti}
    \hat{\nabla}C(K):=\nabla C(K)+ \Delta, 
\end{equation}
\textcolor{black}{where $\Delta \in \mathbb{R}^{\textcolor{blue}{4\times 5}}$ has i.i.d. entries $\Delta_{ij}\sim\mathcal{N}(0,\sigma^2)$, resampled independently at each PGD iteration. We note that this additive noise model is a proxy to study the effect of gradient perturbations; the formal connection between finite noisy data from \eqref{LTI} and gradient estimation errors is established in Section~\ref{sec:MFGD}.}
The PGD algorithm updates the gain ${K}_i$ using \eqref{PGD1} with fixed step size and the gradient estimate from \eqref{arti}. The simulation results are illustrated in Figure \ref{fig:MB}, \textcolor{black}{where lines show the mean and shaded regions indicate one standard deviation, both computed over 100 Monte Carlo runs.
\begin{figure}[h]
    \centering
    \includegraphics[width=0.9\linewidth]{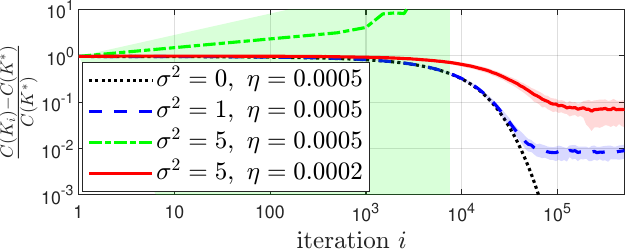}
    \caption{Performance of policy gradient descent with artificially perturbed gradient as in \eqref{arti} for system \eqref{Boeing747} under different \textcolor{blue}{step size $\eta$ and noise $\sigma^2$}, showing mean and standard deviation, over 100 Monte Carlo runs}
    \label{fig:MB}
\end{figure}
When the gradient is exactly known ($\sigma^2=0$, black dotted line), theoretical bounds on the step size guarantee convergence; here $\eta=0.0005$ is shown as an example. When the gradient is affected by small noise ($\sigma^2=1$, blue dashed line), the same step size yields convergence but to a suboptimal value. As the noise level increases ($\sigma^2=5$, green dash-dotted line), the algorithm diverges with large variance. Reducing the step size to $\eta=0.0002$ (red solid line) recovers convergence, but to a suboptimal value.}

\textcolor{black}{These results illustrate that the convergence guarantees obtained in the model-based scenario no longer hold when gradients are affected by noise.} Specifically, the algorithm may converge to suboptimal solutions or even diverge. It is also shown that, by properly tuning the algorithm parameters (such as the step size), convergence can be recovered at the cost of suboptimality. To characterize these fundamental limitations and improve performance, it is paramount to: quantify the uncertainty on gradients and covariances due to finite noisy data; and analyze how this uncertainty propagates in the policy optimization algorithm. To the best of the authors' knowledge, there is no holistic analysis that captures all these aspects and provides guarantees for model-free PG algorithms in the presence of noisy trajectory data. The main technical contribution of this work is to bridge this gap by providing probabilistic guarantees of convergence and suboptimality bounds; closed-form expressions for the algorithm's parameters (e.g., step size, variance reduction) that guarantee converging behavior.

\section{Convergence of Model-free Policy Gradient}\label{sec:MFGD}
\textcolor{black}{In this section, we analyze two variants of model-free policy gradient methods: PGD and NPG, using zeroth-order optimization to estimate the required quantities directly from trajectory data. Section~\ref{secGCE} derives finite-sample error bounds for gradient and covariance estimation under noisy data, and Section~\ref{secGEVR} introduces a variance reduction technique to reduce the required number of rollouts. Building on these results, we establish probabilistic convergence guarantees for PGD (Section~\ref{secMFGD}) and NPG (Section~\ref{secMFNPG}) with adaptive step sizes. Finally, Section~\ref{secE} compares the sample complexity of the proposed methods with and without stochastic noise.}

\subsection{Finite Sample Gradient and Covariance Estimation}\label{secGCE}
 In the model-free setting, zeroth-order optimization techniques \cite{doi:10.1137/1.9780898718768,9186148} can be utilized to approximate the gradient and average covariance matrix using only function evaluations. The procedures for gradient and covariance estimation are outlined in Algorithm \ref{Algo1}.

\begin{algorithm}[h]
  \caption{Gradient and covariance estimation}\label{Algo1}
  \begin{algorithmic}
      \REQUIRE{Gain matrix $K\in \mathcal{S}$, number of rollouts $n$, rollout length $l$, exploration radius $r$}
      \FOR{$k=1,...,n$}
        \STATE{ 1. \textcolor{black}{Generate a sample gain matrix $\bar{K}_k=K+U_k$, where $U_k$ is drawn uniformly at random over matrices of compatible dimensions with Frobenius norm $r$, i.e., $U_k = \frac{r\tilde{U}}{\lVert\tilde{U}\rVert_F}$ with $\tilde{U}$ having i.i.d. $\mathcal{N}(0,1)$ entries};}
        \STATE{ 2. Generate a sample initial state $x_0^{(k)}\sim \mathcal{N}(0,\Sigma_0)$};
        \STATE{ 3. Excite the closed-loop system with $u_t^{(k)}=\bar{K}_kx_t^{(k)}$ for $l$-steps starting from $x_0^{(k)}$, yielding the state sequence $\left\{ x_t^{(k)} \right\}_{t=0}^{l-1}$ originating from \eqref{LTI};}
        \STATE{ 4. Collect the empirical cost estimate $\hat{C}_{\bar{K}_k}:=\frac{1}{l}\sum_{t=0}^{l-1} x_t^{(k)~\top}(Q+\bar{K}^{\top}_kR\bar{K}_k)x_t^{(k)}$ and the empirical covariance matrix $\hat{\Sigma}_k=\frac{1}{l}\sum_{t=0}^{l-1}~x_t^{(k)}x_t^{(k)~\top} $;}
      \ENDFOR
      \RETURN{Covariance estimate $\hat{\Sigma}_{K}:=\frac{1}{n}\sum^{n}_{k=1}\hat{\Sigma}_k$ and gradient estimate $\hat{\nabla}C(K):=\frac{1}{n}\sum^{n}_{k=1}\frac{n_xn_u}{r^2}\hat{C}_{\bar{K}_k}U_k$.}
  \end{algorithmic}
\end{algorithm}
 The empirical cost $\hat{C}_{\bar{K}_k}$ can be computed either using the weighting matrix or directly from performance measurements, without requiring knowledge of $Q$ and $R$, as stated in Remark~\ref{RemarkQR}. We present two theorems analyzing the estimation errors of $\hat{\nabla} C(K)$ and $\hat{\Sigma}_K$ obtained from Algorithm~\ref{Algo1}. 

\textcolor{black}{The estimation error of $\hat{\nabla} C(K)$, denoted $\epsilon$, decomposes into four sources: noisy data ($\epsilon_d$), finite rollout length ($\epsilon_l$), finite number of rollouts in zeroth-order optimization ($\epsilon_n$), and the gap between the gradient of the smoothing function $C_r(K)=\mathbb{E}_{U\sim\mathbb{B}_r}\left[C(K+U)U\right]$ and that of $C(K)$ itself ($\epsilon_r$). The associated failure probability $\delta$ likewise decomposes into three components: $\delta_n$ and $\delta_d$, the concentration-failure probabilities linked to $\epsilon_n$ and $\epsilon_d$ respectively, and $\delta_x$, the probability that the data sample is unbounded.}
 
\begin{Theorem}[Error Bound of $\hat{\nabla}C(K)$]\label{Theorem4}
    Given any tolerance $\epsilon>0$, which can be expressed as $\epsilon=\epsilon_d+\epsilon_l+\epsilon_n+\epsilon_r$, and any probability $\delta\in(0,1)$, which can be expressed as $\delta=1-(1-\delta_d)(1-\delta_n)(1-\delta_x)$ with $\delta_x,\delta_n,\delta_d\in (0,1)$, for a given $K\in \mathcal{S}$,
    the estimated gradient $\hat{\nabla}C(K)$ from Algorithm \ref{Algo1} satisfies:
    \begin{equation*}
    \mathbb{P}\left\{\lVert \hat{\nabla}C(K)- {\nabla}C(K)\rVert\leq \epsilon \right\}\geq 1-\delta,
\end{equation*}
if the parameters $r,l,n$ in Algorithm \ref{Algo1} satisfy:\textcolor{black}{
\begin{subequations}
\begin{align}
r &\leq r_{\max}(C(K),\epsilon_r),\quad l \geq l_{\min}(C(K),\epsilon_l,\epsilon_r),\\
n &\geq n_{\min} := \max\bigl(N_1(C(K),\epsilon_r,\epsilon_n,\delta_n), \notag\\
&\qquad\quad \qquad   \qquad \quad N_2(C(K),\epsilon_r,\epsilon_l,\epsilon_d,\delta_x,\delta_d)\bigr),\label{N2}
\end{align}
\end{subequations}
where \begin{equation*}
\begin{split}
r_{\max} =\mathcal{O}&\left(\min\left(\frac{1}{C(K)},1,\epsilon_r\right)\right);l_{\min}=\mathcal{O}\left(\frac{C(K)^3}{\epsilon_l\epsilon_r}\right);\\
N_1 &=\mathcal{O}\left(\frac{C(K)^2}{\epsilon_r^2\epsilon_n^2}\log\frac{1}{\delta_n}\right); \\
N_2&\textcolor{blue}{=\mathcal{O}\left(\frac{C(K)^{10}}{\epsilon_r^4\epsilon_d^2\epsilon_l^2}\bigg(\log\frac{1}{\delta_x}+\log\frac{1}{\delta_d}\bigg)^2\log\frac{1}{\delta_d}\right).}
\end{split}
\end{equation*}
Detailed expressions for $r_{\max}$, $l_{\min}$, $N_1$ and $N_2$ are given in \eqref{rexpresion}, \eqref{lexpression}, \eqref{Numberofsamples} and \eqref{Nexpression22}, respectively.}
\end{Theorem}
\textcolor{black}{The detailed proof of Theorem \ref{Theorem4} is given in Appendix \ref{ProofTheorem9}. For fixed accuracy $\epsilon_d,\epsilon_l,\epsilon_n,\epsilon_r$ and probability parameters $\delta_d,\delta_n,\delta_x$, given $\Sigma_w=\beta I_{n_x}$ with $\beta>0$, $r_{\max}$ is non-decreasing, while both $l_{\min}$ and $n_{\min}$ are decreasing as $\beta$ decreases.}


\textcolor{black}{Similar to the estimation error of the gradient, the estimation error of $\Sigma_K$, denoted $\epsilon'$, and its associated failure probability $\delta'$ originate from the following sources: $\epsilon'_l$ accounts for the finite rollout length relative to the infinite-horizon average covariance; $\epsilon'_n$ accounts for the noise in the data; and $\epsilon'_r$ accounts for the difference between the average covariance associated with $K$ and $\bar{K}_k$. The failure probability $\delta'$ decomposes into $\delta'_n$, the concentration probability for the noisy data, and $\delta'_x$, the probability that the data sample remains bounded.}
 \begin{Theorem}[Error bound of $\hat{\Sigma}_K$]\label{Theorem5}
    Given an arbitrary tolerance $\epsilon'>0$, which can be expressed as $\epsilon'=\epsilon'_l+\epsilon'_n+\epsilon'_r$, and an arbitrary probability $\delta'\in(0,1)$, which can be expressed as $\delta'=1-(1-\delta'_n)(1-\delta'_x)$ with $\delta'_n,~\delta'_x\in (0,1)$, for a given $K\in \mathcal{S}$, the estimated covariance $\hat{\Sigma}_K$ from Algorithm \ref{Algo1} satisfies
    \begin{equation*}
    \mathbb{P}\left\{\lVert \hat{\Sigma}_K- \Sigma_K \rVert\leq \epsilon' \right\}\geq 1-\delta',
\end{equation*}
    if the parameters $r,l,n$ in Algorithm \ref{Algo1} satisfy:
    \begin{subequations}
       \begin{align}
           r&\leq r'_{\max}(C(K),\epsilon'_r),l\geq l'_{\min}(C(K),\epsilon'_l),\\
           n&\geq n'_{\min}(C(K),\epsilon'_l,\epsilon'_n,\delta'_x,\delta'_n),\label{Nsigma}
       \end{align}
    \end{subequations}
    \textcolor{black}{where 
    \begin{equation*}
\begin{split}
r'_{\max} =\mathcal{O}&\left(\min\left(\frac{1}{C(K)},1,\epsilon'_r\right)\right);\textcolor{blue}{l'_{\min}=\mathcal{O}\left(\frac{C(K)^2}{\epsilon'_l}\right);}\\
n'_{\min} &=\textcolor{blue}{\mathcal{O}\left(\frac{C(K)^4}{{\epsilon'_l}^2{\epsilon'_n}^2}\left(\log\frac{1}{\delta'_x}+\log\frac{1}{\delta'_n}\right)^2\log\frac{1}{\delta'_n}\right).}
\end{split}
\end{equation*}}
Detailed expressions of functions $r'_{\max}$ $l'_{\min}$ and $n'_{\min}$ are given in \eqref{radius2}, \eqref{l2} and \eqref{n22}, respectively.
\end{Theorem}
The proof of Theorem \ref{Theorem5} follows the same steps as Theorem \ref{Theorem4}, except that the gradient of the smoothing function is not involved. \textcolor{black}{A detailed proof can be found in Appendix \ref{PT5}. For fixed accuracy parameters $\epsilon'_l,\epsilon'_n,\epsilon'_r$ and probability parameters $\delta'_n,\delta'_x$, given $\Sigma_w=\beta I_{n_x}$ with $\beta>0$, $r'_{\max}$ is non-decreasing, while both $l'_{\min}$ and $n'_{\min}$ are decreasing as $\beta$ decreases}

The two results above quantitatively analyze the effect of estimation errors caused by stochastic noise acting on \eqref{LTI}. This analysis requires the adoption of new tools and an extension of existing results \cite{pmlr-v80-fazel18a,9254115,10091214} assumed that the trajectory has bounded support. Moreover, while it is generally assumed that all error terms are equal \cite{pmlr-v80-fazel18a}, i.e., $\epsilon_r = \epsilon_n = \epsilon_l$ and $\epsilon_r' = \epsilon_n' = \epsilon_l'$, we present a more structured approach to the error decomposition. By selecting different values for $l$, $r$, and $n$, the same accuracy $\epsilon$ (or $\epsilon'$) can be achieved, but with a more refined breakdown of the contributing error terms. 

\subsection{Gradient Estimation with Variance Reduction}\label{secGEVR}
In the previous section, we employed zeroth-order optimization to estimate the gradient of the cost at the current value of the policy. However, this approach often suffers from high variance, resulting in a slow learning process \cite{6392457}. Using a baseline is a common variance reduction technique in policy gradient methods \cite{6315022,pmlr-v48-mniha16}. In this section, we propose employing the finite-horizon cost function as a baseline. For a state-dependent baseline function $b(x)$, the estimated gradient with variance reduction $\hat{\nabla} C_v(K)$ is expressed as
\begin{equation*}
    \hat{\nabla} C_v(K):=\frac{1}{n}\sum^{n}_{k=1}\frac{n_xn_u}{r^2}\big(\hat{C}_{\bar{K}_k}-b(x_0^{(k)})\big)U_k,
\end{equation*}
where $U_k,x_0^{(k)},\hat{C}_{\bar{K}_k},r$ are the same as those defined in Algorithm \ref{Algo1}. The estimated gradient $\hat{\nabla} C_v(K)$ satisfies:
\begin{equation*}
    \begin{split}
        \mathbb{E}_{x_0^{(k)},U_k}&[\hat{\nabla} C_v(K)]=\mathbb{E}_{x_0^{(k)},U_k}[\hat{\nabla} C(K)]\\
        &-\frac{1}{n}\sum^{n}_{k=1}\frac{n_xn_u}{r^2}\mathbb{E}_{x_0^{(k)},U_k}[ b(x_0^{(k)})U_k].
    \end{split}
\end{equation*}
Since $U_k$ and $x_0^{(k)}$ are independent (as described in Algorithm \ref{Algo1}), then $\mathbb{E}_{x_0^{(k)},U_k}[\hat{\nabla} C_v(K)]=\mathbb{E}_{x_0^{(k)},U_k}[\hat{\nabla} C(K)].$ Defining $Z_k^V := (\hat{C}_{\bar{K}_k} - b(x_0^{(k)}))U_k$, the trace of the covariance matrix of $\textcolor{black}{\mathrm{vec}}(Z_k^V)$, i.e. $\mathrm{Tr}(\mathrm{Var}(\textcolor{black}{\mathrm{vec}}(Z_k^V)))$, denoted by $z_c$, is given by
\begin{equation*}
   \begin{split}
        z_c=&\mathbb{E}_{x_0^{(k)},U_k}\left[\left\lVert Z_k^V \right\rVert_F^2 \right] -\left\lVert\mathbb{E}_{x_0^{(k)},U_k}\left[ Z_k^V \right]\right\rVert_F^2.
   \end{split}
\end{equation*}
Next, we determine the optimal function $b^*$, which minimizes the trace of the variance matrix above:
\begin{equation}\label{VR2}
\begin{split}    
z_c=\mathbb{E}
_{{\substack{x_0^{(k)}, U_k}}}\bigg[\big\lVert (\hat{C}_{\bar{K}_k} -b(x_0^{(k)}))U_k\big\rVert_F^2 \bigg]\\
-\left\lVert\mathbb{E}_{x_0^{(k)},U_k}\left[ (\hat{C}_{\bar{K}_k} -b(x_0^{(k)}))U_k \right]\right\rVert_F^2.  
\end{split}
\end{equation}
For any state-dependent baseline function $b$, since $U_k$ is i.i.d and zero mean, we have: $\mathbb{E}_{x_0^{(k)},U_k}\left[ b(x_0^{(k)})U_k\right]=0.$ Consequently, the second term in \eqref{VR2} becomes independent of the choice of $b$, leaving only the first term in \eqref{VR2} to be minimized. To do so, the baseline function should be chosen to reduce the mean square error of $\hat{C}_{\bar{K}_k}$. Thus, the optimal baseline $b^*$ is given by:
\begin{equation}\label{eqbaseline}
    b^*(x_0)=\frac{1}{l}\sum_{t=0}^{l-1} \mathbb{E}_{w_t,U_k} \left[ x_t^{~\top}Q_{K+U_k}x_t \right].
\end{equation}
Since $U_k$, which was also already defined in Algorithm \ref{Algo1}, enters nonlinearly in \eqref{eqbaseline}, it is hard to compute the expectation and thus to obtain a closed-form expression for $b^*(x_0)$. While it is possible to estimate it directly from data, the lack of a closed-form solution makes it challenging to establish \textcolor{black}{sample complexity analysis}. \textcolor{black}{We therefore propose the following tractable approximation, obtained by setting $x_0=0$ and removing the dependence on $U_k$:}
\begin{equation}\label{truebaseline}
    b_s=\frac{1}{l}\sum_{t=0}^{l-1} \mathbb{E}_{w_t} \left[ x_t^{~\top}Q_Kx_t \right],~\mathrm{with} ~x_0=0.
\end{equation}
\textcolor{black}{This choice is motivated by two considerations: it yields a tractable closed-form expression for the baseline, enabling an explicit characterization of the sample complexity analysis established in Theorem~\ref{TheoremVariancereduction}, and it avoids re-estimating the baseline inside the gradient estimation loop, since $b_s$ depends only on the current gain $K$ and not on the initial state $x_0^{(k)}$, thus saving additional rollouts at each iteration. }

Algorithm \ref{AlgoV} summarizes the procedure used to estimate the baseline function from noisy data. \textcolor{black}{In Theorem \ref{TheoremVariancereduction}, we will show that this approximated baseline function can be estimated with high accuracy using a finite number of samples, as characterized later in \eqref{bs}.}
\begin{algorithm}[H]
  \caption{Baseline estimation $\hat{b}_s$ for variance reduction}\label{AlgoV}
  \begin{algorithmic}
      \REQUIRE{Gain matrix $K\in \mathcal{S}$, number of rollouts to estimate the baseline function $n_v$, rollout length $l$ }
      \FOR{$k=1,...,n_v$}
        \STATE{ 1. Excite the closed-loop system with $u_t=Kx_t$ for $l$-steps starting from $x_0^{(k)}=0$, yielding the state sequence $\left\{ x_t^{(k)} \right\}_{t=0}^{l-1}$;}
        \STATE{ 2. Collect the empirical finite-horizon cost estimate $\hat{C}^V_k:=\frac{1}{l}\sum_{t=0}^{l-1} x_t^{(k)~\top}(Q+{K}^{\top}R{K})x_t^{(k)}$;}
      \ENDFOR
      \RETURN{Baseline function estimate $\hat{b}_s:=\frac{1}{n_v}\sum^{n_v}_{k=1}\hat{C}^V_k$.}
  \end{algorithmic}
\end{algorithm}

Using the estimated baseline function $\hat{b}_s$, the gradient estimation algorithm is given in Algorithm \ref{AlgoVV}. 

\begin{algorithm}[H]
  \caption{Gradient estimation with variance reduction}\label{AlgoVV}
  \begin{algorithmic}
      \REQUIRE{Gain matrix $K\in \mathcal{S}$, number of rollouts $n_b$, rollout length $l$, exploration radius $r$ }
      \STATE{Estimate the baseline function $\hat{b}_s$ using Algorithm \ref{AlgoV}}
      \FOR{$k=1,...,n_b$}
             \STATE{ 1. Generate a sample initial state $x_0^{(k)}$;}
        \STATE{ 2. Generate a sample gain matrix $\bar{K}_k=K+U_k$, where $U_k$ is drawn uniformly at random over matrices of compatible dimensions with Frobenius norm $r$; }
        \STATE{ 3. Excite the closed-loop system with $u_t^{(k)}=\bar{K}_kx_t^{(k)}$ for $l$-steps starting from $x_0^{(k)}$, yielding the state sequence $\left\{ x_t^{(k)} \right\}_{t=0}^{l-1}$;}
        \STATE{ 4. Collect the empirical finite-horizon cost estimate $\hat{C}_{\bar{K}_k}:=\frac{1}{l}\sum_{t=0}^{l-1} x_t^{(k)~\top}(Q+\bar{K}^{\top}_kR\bar{K}_k)x_t^{(k)}$}
      \ENDFOR
      \RETURN{$\hat{\nabla}C(K):=\frac{1}{n_b}\sum^{n_b}_{k=1}\frac{n_xn_u}{r^2}(\hat{C}_{\bar{K}_k}-\hat{b}_s)U_k$}
  \end{algorithmic}
\end{algorithm}

Compared with Algorithm \ref{Algo1}, Algorithm \ref{AlgoVV} introduces an additional step involving the baseline function to reduce the variance of the estimated gradient. \textcolor{black}{The following theorem shows that this modification does not increase the sample complexity required to achieve a given gradient accuracy with high probability, where the failure probability $\delta_v$ captures the concentration of the sample. \textcolor{blue}{For ease of comparison, we choose $\epsilon_n=\epsilon_d$ and $\delta_n=\delta_d$, so that the required number of samples $n_{\min}$ in \eqref{N2} from Theorem~\ref{Theorem4} is dominated by the term $N_2$, i.e., $n_{\min}=N_2$; this simplifies the interpretation of the results in the theorem that follows.}}

\begin{Theorem}[Variance Reduction Guarantee]\label{TheoremVariancereduction}
    Given the same tolerance $\epsilon=\epsilon_d+\epsilon_l+\epsilon_n+\epsilon_r$ with $\epsilon_d=\epsilon_n$ and probability $\delta=1-(1-\delta_d)(1-\delta_n)(1-\delta_x)$ with $\delta_n=\delta_d$ introduced in Theorem \ref{Theorem4}, for a given $K\in \mathcal{S}$,
    then the estimated gradient $\hat{\nabla}C(K)$ from Algorithm \ref{AlgoVV} enjoys the following bound:
    \begin{equation*}
    \mathbb{P}\left\{\lVert \hat{\nabla}C(K)- {\nabla}C(K)\rVert\leq \epsilon \right\}\geq 1-\delta,
\end{equation*}
    if $n_b,l$ and $r$ in Algorithm \ref{AlgoVV} satisfy: 
    \begin{subequations}
       \begin{align}
            r&\leq r_{\max}(C(K),\epsilon_r),l\geq l_{\min}(C(K),\epsilon_l,\epsilon_r),\label{1aaa}\\
            n_b& \textcolor{blue}{\geq N_b(\hat{b}_s)\label{1bbb}=\mathcal{O}\bigg(\frac{1}{\epsilon_d^2}\bigg(\frac{C(K)^5}{\epsilon_r^2\epsilon_l}-\frac{b_s}{\epsilon_r}}\\&\textcolor{blue}{+\frac{|\hat{b}_s-b_s|}{\epsilon_r}\bigg)^2\bigg(\log\frac{1}{\delta_x}+\log\frac{1}{\delta_d}\bigg)^2\log\frac{1}{\delta_d}\bigg)}\notag.
        \end{align}
    \end{subequations}
The detailed expression of $N_b$ is given in \eqref{n3expression} in Appendix \ref{DetailedExpression} and $r_{\max}$ and $l_{\min}$ were introduced in Theorem \ref{Theorem4}.  
    
    Moreover, let the desired probability $\delta_v\in(0,1)$ and let $\epsilon_v \in (0,1)$ denote the relative estimation accuracy for the baseline function \textcolor{black}{i.e. $|\hat{b}_s-b_s|\leq \epsilon_vb_s$}. If the number of rollouts $n_v$ for baseline function estimation (as defined in Algorithm \ref{AlgoV}) satisfies 
    \textcolor{blue}{\begin{equation}\label{bs}
    \begin{split}
        n_v\geq \tilde{n}_{\min}&(C(K),l,{\delta}_v,\epsilon_v)=\mathcal{O}\left(\frac{C(K)^{2}l}{\epsilon_v^2b_s^2}\log\frac{1}{{\delta}_v}\right),
    \end{split}  
    \end{equation}}
    then:
    \begin{equation*}\label{1ccc}
        \mathbb{P}\left\{n_{\min} \geq  N_b(\hat{b}_s) \right\}\geq 1-\delta_v,
    \end{equation*}
    where $n_{\min}$ was first introduced in \eqref{N2} in Theorem \ref{Theorem4} and the detailed expression of $\tilde{n}_{\min}$ is provided in \eqref{tilden} in Appendix \ref{ProofVariance}. 
\end{Theorem}

 The proof of Theorem~\ref{TheoremVariancereduction} is given in Appendix~\ref{ProofVariance}. Theorem~\ref{TheoremVariancereduction} shows that the exploration radius $r_{\max}$ and rollout length $l_{\min}$ are unchanged by variance reduction, while the required number of rollouts for gradient estimation is provably no greater than $n_{\min}$ when the baseline is estimated with sufficient accuracy. 
 \textcolor{blue}{The expression for $N_b$ in \eqref{1bbb} has a clear structure compared to $N_2$ in \eqref{N2}, it adds two terms, $-\frac{b_s}{\epsilon_r}$ reflecting the gain from variance reduction, and $\frac{|\hat{b}_s-b_s|}{\epsilon_r}$ reflecting the loss from estimation error of the baseline function.}
However, estimating the baseline requires an additional $\tilde{n}_{\min}$ rollouts. The following corollary shows that the net effect is still favorable.
\begin{Corollary}[Total Sample Complexity]\label{Coro1}
Under the notation of Theorem~\ref{TheoremVariancereduction}, choose $\delta_d={\delta}_v$. Suppose $\epsilon_v\in(0,1)$ satisfies $\epsilon_v\geq\frac{\epsilon_d}{b_s}$ and \textcolor{blue}{$\epsilon_v\leq\min\{\frac{1}{b_s},1-\frac{\beta r^2}{64n_x^2n_u^2b_s}\}$ with $\beta:=\frac{\lambda_{n_x}(Q+K^\top RK)^2\lambda_1(\Sigma_w)}{\lambda_1(Q)}\cdot\frac{\lambda_1(\Sigma_w)+\lambda_{n_x}(\Sigma_0)}{\lambda_1(\Sigma_w)}$}. Then
\begin{equation*}
\mathbb{P}\left\{n_{\min}> N_b(\hat{b}_s)+\tilde{n}_{\min}(C(K),l,{\delta}_v,\epsilon_v)\right\}\geq 1-\delta_v,
\end{equation*}
where $n_{\min}$ was introduced in Theorem \ref{Theorem4}; $N_b$ and $\tilde{n}_{\min}$ come from Theorem \ref{TheoremVariancereduction}.
\end{Corollary}

The proof follows from comparing the closed-form expressions of $n_{\min}$, $N_b$, and $\tilde{n}_{\min}$; see Appendix \ref{Proofcoro}. \textcolor{black}{Corollary~\ref{Coro1} establishes that the total sample complexity of Algorithm~\ref{AlgoVV}, including the $\tilde{n}_{\min}$ rollouts for baseline estimation, strictly smaller than that of Algorithm~\ref{Algo1}. The condition $\epsilon_v\geq\frac{\epsilon_d}{b_s}$ ensures that the baseline need not be estimated more accurately than the gradient itself, avoiding unnecessary computational cost. The upper bound condition on $\epsilon_v$ ensures that $N_b(\hat{b}_s)$ is sufficiently smaller than $n_{\min}$ to offset the additional cost $\tilde{n}_{\min}$. Both conditions are most easily satisfied when $\epsilon_d$ and $r$ are small, i.e., precisely the high-accuracy regime where variance reduction is most beneficial.}

\textcolor{black}{To the best of our knowledge, this is the first work to formally analyze variance reduction for LQR using a baseline function estimated from noisy data, providing provable improvements in sample complexity guarantees.} 
\subsection{Model-free Policy Gradient Descent}\label{secMFGD}
Building on the gradient uncertainty quantified in the previous section, we now proceed to study the convergence of the model-free PGD algorithm, which is also summarized in Algorithm \ref{Algo2}. \textcolor{black}{In the rest of this paper, $\hat{K}_i$ denotes the gain at the $i$-th iteration updated using gradient estimates.}
\begin{algorithm}[H]
  \caption{Model-free PGD with adaptive step size}\label{Algo2}
  \begin{algorithmic}
      \REQUIRE{An initial stabilizing gain matrix $\hat{K}_0$, desired accuracy $\epsilon$ and probability $\delta$}.
      \FOR{$i=\textcolor{black}{0},...,\infty$}
      \STATE {1. Compute the required rollouts $n_i$, exploration radius $r_i$ and rollout length $l_i$ based on Theorem \ref{Theorem4} or \ref{TheoremVariancereduction} to achieve the accuracy stated in Theorem \ref{MFPGV}.}
    \STATE {2. Use Algorithm \ref{Algo1} or \ref{AlgoVV} to estimate the gradient $\hat{\nabla} C(\hat{K}_i)$.}
      \STATE{ 3. Update the \textcolor{black}{gain} as $\hat{K}_{i+1}=\hat{K}_i-\eta_i \hat{\nabla} C(\hat{K}_i)$, with $\eta_i\leq h_\mathrm{PGD}(C(\hat{K}_i))$.}
      \ENDFOR 
  \end{algorithmic}
\end{algorithm}
\begin{Theorem}[Model-free PGD] \label{MFPGV}
    Suppose the initial $\hat{K}_0\in \mathcal{S}$, and consider the policy gradient descent with adaptive step size:
    \begin{equation*}\label{GDS}
    \hat{K}_{i+1}=\hat{K}_i-\eta_i \hat{\nabla} C(\hat{K}_i), \quad \forall i \in \mathbb{Z}_+,
\end{equation*}
where $\hat{\nabla} C(\hat{K}_i)$ is the gradient estimate from Algorithm \ref{Algo1} \textcolor{black}{or Algorithm \ref{AlgoVV}} and $0 <\eta_i \leq h_{\mathrm{PGD}}(C(\hat{K}_i))$, where the function $h_{\mathrm{PGD}}$ was introduced in Theorem \ref{PGTheorem}. 
Given any accuracy $\epsilon >0 $,  and $\sigma\in (0,1)$, define ${\eta_\mathrm{PGD}}:=\inf\limits_{i}\eta_i$ and the number of iterations $n_{\mathrm{PGD}}$:
\begin{equation*}
    n_{\mathrm{PGD}}\textcolor{black}{\geq}\frac{\lVert \Sigma_{K^*}\rVert}{2(1-\sigma)\eta_\mathrm{PGD} \lambda_1(R)\lambda^2_1(\Sigma_w)}\log \frac{C(\hat{K}_{0})-C(K^*)}{\epsilon}.
\end{equation*}
Given any probability $\delta \in (0,1)$ satisfying $\delta n_{\mathrm{PGD}}\in (0,1)$, assume that for all $i\in [0,n_{\mathrm{PGD}}]$, the estimation error of the gradient $\hat{\nabla} C(\hat{K}_i)$ satisfies:
\begin{equation*}
    \mathbb{P}\bigg\{\lVert \hat{\nabla}C(\hat{K}_i)- {\nabla}C(\hat{K}_i)\rVert\leq \frac{2\sigma\epsilon\lambda_1{(R)} \lambda_1^2(\Sigma_w)}{h_{C}(C(\hat{K}_i))\lVert \Sigma_{K^*}\rVert} \bigg\}\geq 1-\delta,
\end{equation*}
where $h_C$ was introduced in Lemma \ref{Lipschitz}. Then for any $C(\hat{K}_i)\geq C(K^*)+\epsilon$, the following inequality holds:
\begin{equation*}
        \mathbb{P}\left\{C(\hat{K}_{i+1})-C(K^*) \leq  \gamma_i (C(\hat{K}_{i})-C(K^*))\right\}\geq 1-\delta,
\end{equation*}
where $\gamma_i:= 1-(1-\sigma)\frac{2\eta_i \lambda_1{(R)} \lambda_1^2(\Sigma_w)}{\lVert \Sigma_{K^*}\rVert}$ and $\gamma_i < 1, \forall i \in \mathbb{Z}_+$. \\
As a result, the policy gradient descent method enjoys the following performance bound:
\begin{equation*}
    \mathbb{P}\bigg\{ \min_{i\in [0,n_{\mathrm{PGD}}]}C(\hat{K}_i)-C(K^*)  \leq \epsilon \bigg\}\geq 1-\delta n_{\mathrm{PGD}}.
\end{equation*}
\end{Theorem}
The proof of Theorem \ref{MFPGV} is given in Appendix \ref{ProofMFPG}

To achieve the desired accuracy of the gradient estimates stated in Theorem \ref{MFPGV}, the corresponding values of $l_i, r_i, n_i$ should be selected based on Theorem \ref{Theorem4}. \textcolor{blue}{Given the desired suboptimality gap $\epsilon$, we choose $\epsilon_d=\epsilon_l=\epsilon_n=\epsilon_r=\epsilon$ for Theorem \ref{Theorem4}, the total sample complexity of Algorithm \ref{Algo2} is given by $\mathcal{O}({\epsilon^{-8}}\ln \epsilon^{-1})$}.  Theorem \ref{MFPGV} ensures that the cost will converge to the optimal solution with a predefined accuracy gap $\epsilon$, meaning $C(K^*)+\epsilon$ will be reached with a certain probability. However, due to the estimation error in the gradient, further improvements in the cost stop when the realized cost satisfies $C(\hat{K}_i) \leq C(K^*)+\epsilon$. To achieve higher accuracy, it is necessary to reduce the gradient estimation error. Nevertheless, the algorithm can only converge to the optimal when the gradient estimation is exact, which is not possible with a finite number of rollouts. The interpretation of the $\sigma$ is given later in Remark \ref{mbmfsigma}. Regarding the step size $\eta_i$ stated in Theorem \ref{MFPGV}, from the expression of $h_\mathrm{PGD}$ given in \eqref{pG2} we observe that as the noise level $\Sigma_w$ increases, the step size $\eta_i$ decreases, reflecting the observation analyzed in Section \ref{sec:MBGD}\ref{simu24}.

In Algorithm \ref{Algo2}, the required rollouts $n_i$, exploration radius $r_i$ and rollout length $l_i$ are determined adaptively to the cost $C(\hat{K}_i)$, i.e., they are computed online inside the for-loop. This design is aligned with Theorem \ref{MFPGV}, which shows that the cost function $C(\hat{K}_i)$ decreases probabilistically with each iteration. As the cost $C(\hat{K}_i)$ decreases, this leads to a reduction in the required rollouts $n_i$ and rollout length $l_i$, while the exploration radius $r_i$ increases. Alternatively, the algorithm can be modified to compute these parameters offline, where for all iterations $i$, $n_i,l_i,r_i$ are calculated once based on the initial cost $C(\hat{K}_0)$.

\textcolor{black}{
\begin{Remark}
The bounds on the algorithm parameters $r_{\max}$, $l_{\min}$, $n_{\min}$, and $\eta_i$ in Theorem~\ref{MFPGV} depend on system quantities such as $\|A\|$ and $\|B\|$, which are not available in the model-free setting. This dependence can be found in several of the works on zeroth-order PG methods, e.g. \cite{pmlr-v80-fazel18a,9254115,carnevale2025datadrivenlqrfinitetimeexperiments,10091214}. These bounds should therefore be interpreted as complexity estimates that quantify how the algorithmic parameters scale with the desired accuracy and the problem noise level to guarantee convergence. Rather than merely identifying qualitative trends, the analysis provides explicit rates that characterize the strength of these dependencies, thereby offering practical guidance for parameter selection. 
\end{Remark}}
\subsection{Model-free Natural Policy Gradient}\label{secMFNPG}
We now turn to analyzing the convergence of the NPG method, which is also summarized in Algorithm \ref{Algo3}.

\begin{algorithm}[H]
  \caption{Model-free NPG with adaptive step size}\label{Algo3}
  \begin{algorithmic}
      \REQUIRE{An initial stabilizing gain matrix $\hat{K}_0$, desired accuracy $\epsilon$ and probability $\delta$}.
      \FOR{$i=0,...,\infty$}
    \STATE {1. Compute the required rollouts $n_i$, exploration radius $r_i$ and rollout length $l_i$ based on Theorems \ref{Theorem4} and \ref{Theorem5} to achieve the accuracy stated in Theorem \ref{MFNPGV}.} 
    \STATE {2. Use Algorithm \ref{Algo1} to estimate the gradient $\hat{\nabla}C(\hat{K}_i)$ and covariance $\hat{\Sigma}_{\hat{K}_i}$.}
      \STATE{ 3. Update the \textcolor{black}{gain} as $\hat{K}_{i+1}=\hat{K}_i-\eta_i \hat{\nabla} C(\hat{K}_i)\hat{\Sigma}^{-1}_{\hat{K}_i}$ with \textcolor{black}{$\eta_i\leq \frac{1}{2\lVert R\rVert+\frac{2\lVert B \rVert^2 C(\hat{K}_i)}{\lambda_1(\Sigma_w)}}$}.}
      \ENDFOR 
  \end{algorithmic}
\end{algorithm}

\begin{Theorem}[Model-free NPG] \label{MFNPGV}
    Suppose the initial $\hat{K}_0\in \mathcal{S}$, and consider natural policy gradient with adaptive step size:   
    \begin{equation}\label{theorem4.7update}
    \hat{K}_{i+1}=\hat{K}_i-\eta_i \hat{\nabla}C(\hat{K}_i)\hat{\Sigma}_{\hat{K}_i}^{-1},\quad \forall i \in \mathbb{Z}_+,
\end{equation}
where $\hat{\nabla} C(\hat{K}_i)$ and $\hat{\Sigma}_{\hat{K}_i}$ are the gradient and covariance estimates from Algorithm \ref{Algo1} and \textcolor{black}{$0<\eta_i\leq \frac{1}{2\lVert R\rVert+\frac{2\lVert B \rVert^2 C(\hat{K}_i)}{\lambda_1(\Sigma_w)}}$}. Given any accuracy $\epsilon> 0$ and $\sigma\in (0,1)$, 
defining ${\eta_\mathrm{NPG}}:=\inf\limits_{i}\eta_i$ and the number of iterations $n_\mathrm{NPG}$:
\textcolor{black}{\begin{equation*}
    n_\mathrm{NPG} \geq \frac{\lVert \Sigma_{K^*}\rVert}{2(1-\sigma)\eta_{\mathrm{NPG}}\lambda_1(R)\lambda_1(\Sigma_w)}\log \frac{C(\hat{K}_{0})-C(K^*)}{\epsilon}.
\end{equation*}}
Given any probability $\delta \in (0,1)$ satisfying $\delta n_{\mathrm{NPG}}\in (0,1)$, assume that for all $i\in [0,n_{\mathrm{NPG}}]$, the estimation error of the gradient $\hat{\nabla} C(\hat{K}_i)$ and the covariance $\hat{\Sigma}_{\hat{K}_i}$ satisfy:
\begin{subequations}
\begin{align}
    \mathbb{P}&\left\{\lVert \hat{\nabla} C(\hat{K}_i) -{\nabla} C(\hat{K}_i)\rVert \leq\Delta_i^\nabla  \right\}\geq 1-\frac{\delta}{2},\label{1a}\\
    \mathbb{P}&\left\{\lVert\hat{\Sigma}_{\hat{K}_i} -{\Sigma}_{\hat{K}_i} \rVert \leq \textcolor{black}{\Delta_i^{\Sigma}}\right\}\geq {1-\frac{\delta}{2}},\label{1b}
\end{align}
\end{subequations}
where $\Delta_i^\Sigma:=\frac{\sigma\epsilon \lambda_1{(R)} \lambda_1(\Sigma_w)}{4h_{{\textcolor{blue}{C}}}(C(\hat{K}_i))\textcolor{blue}{\lVert \Sigma_{K^*}\rVert}{b_{\nabla}(C(\hat{K}_i))}}$; $\Delta_i^\nabla:=\frac{\sigma\epsilon\lambda_1{(R)} \lambda_1^2(\Sigma_w)}{2h_{{\textcolor{blue}{C}}}(C(\hat{K}_i))\textcolor{blue}{\lVert \Sigma_{K^*}\rVert}}$; $h_{\textcolor{blue}{C}}$ was introduced in Lemma \ref{Lipschitz} and $b_{\nabla}$ was introduced in Lemma \ref{bound}. Then for any $C(\hat{K}_i)\geq C(K^*)+\epsilon$, the following inequality holds:
\begin{equation*}
        \mathbb{P}\left\{    C(\hat{K}_{i+1})-C(K^*) \leq  \kappa_i (C(\hat{K}_{i})-C(K^*))\right\}\geq 1-\delta,
\end{equation*}
where $\kappa_i\textcolor{blue}{:=} 1-(1-\sigma)\frac{2\eta_i \lambda_1{(R)} \lambda_1(\Sigma_w)}{\textcolor{blue}{\lVert \Sigma_{K^*}\rVert}}$ and $\kappa_i<1,\forall i \in \mathbb{Z}_+$. As a result, the natural policy gradient method enjoys the following performance bound:
\begin{equation*}
    \mathbb{P}\bigg\{ \min_{i\in[0,n_\mathrm{NPG}]}C(\hat{K}_i)-C(K^*)  \leq \epsilon \bigg\}\geq 1-\delta{n_\mathrm{NPG}}.
\end{equation*}
\end{Theorem}
The proof of Theorem \ref{MFNPGV} is provided in Appendix \ref{ProofMFNPGV}.

Suitable values of $l^{\nabla}_i,r^{\nabla}_i,n^{\nabla}_i$ to satisfy the requirement for the gradient estimation error, as indicated in \eqref{1a}, can be selected based on Theorem \ref{MFNPGV}. Similarly, we can select $l^{\Sigma}_i,r^{\Sigma}_i,n^{\Sigma}_i$ to satisfy the requirement for the covariance estimation error, as described in \eqref{1b} from Theorem \ref{MFNPGV}. To ensure both requirements are met, we can then set: $n_i \geq\max\{n^{\nabla}_i,n^{\Sigma}_i\},l_i\geq \max\{l^{\nabla}_i,l^{\Sigma}_i\}, r_i\leq \min\{r^{\nabla}_i,r^{\Sigma}_i\}$. These choices guarantee that the estimates of both the gradient and the covariance matrix satisfy their respective accuracy requirements. \textcolor{blue}{Given the desired suboptimality gap $\epsilon$, we choose $\epsilon_d=\epsilon_l=\epsilon_n=\epsilon_r=\epsilon$ for Theorem \ref{Theorem4} and $\epsilon'_l=\epsilon'_n=\epsilon'_r=\epsilon$ for Theorem \ref{Theorem5}, the total sample complexity of Algorithm \ref{Algo3} is given by $\mathcal{O}({\epsilon^{-8}}\ln \epsilon^{-1})$.}

\begin{Remark}\label{mbmfsigma}
The convergence rate in the model-free setting is typically slower than in the model-based case by a factor of ($1-\sigma$). This reduction is due to estimation errors in the gradient and covariance matrices, which are inherent in model-free methods as they rely on data-driven estimates rather than exact model information. When the estimated gradient closely matches the true gradient, the model-free policy gradient approaches the same convergence rate as the model-based policy gradient, corresponding to $\sigma \to0$. However, if the estimation error becomes large, such that $\sigma\to 1$, further policy improvements become impossible, and the algorithm can no longer guarantee the contraction at each iteration. This effect of gradient estimation error on the convergence rate and suboptimality gap aligns with our observations in Figure \ref{fig:MB}, as illustrated by the black dotted line, blue dashed line, and green dash-dotted line. 
\end{Remark}

\subsection{Comparison of Model-free PG Methods for Noise and Noise-free Cases}\label{secE}
In Algorithm \ref{Algo1}, three key quantities, i.e. exploration radius $r$, rollout length $l$, and the required rollouts $n$, play a central role in determining the estimation errors on gradient and covariance estimation. The presence of stochastic noise acting on the system \eqref{LTI} during data collection influences  these quantities in the following ways:
\begin{itemize}
    \item \textbf{Exploration radius:} The selection of the exploration radius is fully determined by the desired accuracy and the specific model parameters. The choice of $r$ directly influences the estimation error by affecting the bias term $U_k$, which represents the difference between quantities $\lVert \nabla C(K) \rVert$ and $\lVert \nabla C(K+U_k) \rVert$, as well as $\lVert \Sigma_K \rVert$ and $\lVert \Sigma_{K+U_k} \rVert$. These differences are crucial in the error analysis, as detailed in Lemma \ref{Lemma12} and Appendix \ref{PT5} respectively. The estimation error of the gradient is discussed in detail in \eqref{rexpresion} for the noise case and in \cite[Lemma 27]{pmlr-v80-fazel18a} for the noise-free case. Similarly, the estimation error of the covariance is analyzed in \eqref{radius2} for the noise case and in \cite[Theorem 30]{pmlr-v80-fazel18a} for the noise-free case. Due to the differences in cost functions, for the noisy case, $r$ is determined by $\Sigma_w$ whereas for the noise-free case, it is determined by $\Sigma_0$.
    \item \textbf{Rollout length:} The rollout length $l$ is different for the noise-free case and the noise case. In both scenarios, the rollout length plays a crucial role in estimating the gradient and covariance, whose true values are defined over an infinite horizon. However, finite approximations must be used. In the noise-free case, as described in \cite[Lemma 23]{pmlr-v80-fazel18a}, the rollout length is determined by the desired accuracy, model parameters, and initial covariance $\Sigma_0$. In the presence of noise, as discussed in Lemma \ref{Lemmafinitness}, the rollout length is influenced not only by $\Sigma_0$, but also by the noise covariance $\Sigma_w$. Therefore, the rollout length in the noisy case must account for both the system's initial conditions and noise disturbances. These relationships are quantitatively explored in Lemma \ref{Lemmafinitness} by \eqref{l11} and \eqref{ll1}.
    \item \textcolor{black}{\textbf{Required rollouts:} The most significant difference between the noisy and noise-free cases comes from applying matrix concentration inequalities to trajectories consisting of noisy data. In the noise-free case, boundedness of the data is guaranteed by an upper bound on the initial state \cite[Lemmas 27, 29]{pmlr-v80-fazel18a}. In contrast, with unbounded additive Gaussian noise, the state remains bounded only with a certain probability (Lemma~\ref{boundofstate}), introducing an additional failure probability $\delta_x,\delta'_x$ into the analysis. The probabilistic sample bound depends on the rollout length $l$, and since the variance bound requires squaring this quantity, the overall sample complexity increases significantly compared to the noise-free case.}

\textcolor{black}{To formally quantify this, \textcolor{blue}{we assume $\bar{\epsilon}=\epsilon_d=\epsilon_l=\epsilon_n=\epsilon_r$, $\bar{\delta}=\delta_d=\delta_n=\delta_x$, and similarly $\bar{\epsilon}'=\epsilon'_l=\epsilon'_n=\epsilon'_r$, $\bar{\delta}'=\delta'_n=\delta'_x$ as in \cite{pmlr-v80-fazel18a}}. The sample complexity from \cite{pmlr-v80-fazel18a} is $n_{\min}^{\nabla}=\mathcal{O}\left(\frac{1}{\bar{\epsilon}^4}\log\frac{1}{\bar{\delta}}\right)$ for gradient estimation and $n^{\Sigma}_{\min}=\mathcal{O}\left(\frac{1}{\bar{\epsilon}'^2}\log\frac{1}{\bar{\delta}'}\right)$ for covariance estimation, while ours, determined by $N_2$ in \eqref{N2} and $n'_{\min}$ in \eqref{Nsigma}, are $n_{\min}=\mathcal{O}\left(\frac{1}{\bar{\epsilon}^8}\log^3\frac{1}{\bar{\delta}}\right)$ and $n'_{\min}=\mathcal{O}\left(\frac{1}{\bar{\epsilon}'^4}\log^3\frac{1}{\bar{\delta}'}\right)$, respectively. In both cases, the additional factors reflect the cost of handling unbounded additive Gaussian noise via probabilistic sample boundedness, which is absent in the noise-free setting of \cite{pmlr-v80-fazel18a}. Note that these comparisons ignore dependence on system parameters such as $C(K)$, $\lambda_1(\Sigma_w)$, and $\lambda_1(\Sigma_0)$, which differ in structure between the two settings.}
\end{itemize}

\section{Simulation Results}
\label{sec:Simulation}
In this section, we use numerical simulations \footnote{The Matlab codes used to generate these results (in Section \ref{sec:MBGD}\ref{simu24} and Section \ref{sec:Simulation}) are accessible from the repository: {https://github.com/col-tasas/2025-PGforLQRwithStochastic}} to demonstrate the benefits of incorporating the variance reduction technique and adaptive step sizes in model-free policy gradient algorithms and the overall effect of noise on them. The matrices $(A,B,Q,R)$ and $\hat{K}_0$ are the same as \textcolor{black}{the longitudinal dynamics of the Boeing 747
aircraft in by \eqref{Boeing747} in Section \ref{sec:MBGD}\ref{simu24}}. The rollout length, exploration radius, and number of rollouts are set to \textcolor{black}{$l=20,~r=0.01,~n=500$}, respectively. The simulation results are obtained from a Monte Carlo simulation over \textcolor{black}{$100$} data samples.
\textcolor{black}{\subsection{Gradient Estimation}\label{simGE}
Figure~\ref{fig:fig5} plots the normalized gradient estimation error at the initial gain $\hat{K}_0$, comparing Algorithm~\ref{Algo1} and Algorithm~\ref{AlgoVV} across different noise levels. We normalize the estimation error by $\lVert E_{\hat{K}_0} \rVert$ instead of $\lVert \nabla C(\hat{K}_0)\rVert$ to remove the dependence of the normalization factor on $\Sigma_w$ (see \eqref{cSigmaK}).
The total number of rollouts is fixed at $n=500$; for variance reduction, $n_v=100$ rollouts are used to estimate the baseline function (Algorithm \ref{AlgoV}) and the remaining $n_b=400$ for gradient estimation (Algorithm \ref{AlgoVV}), keeping the total sample budget identical.
\begin{figure}[H]
    \centering
    \includegraphics[width=1\linewidth]{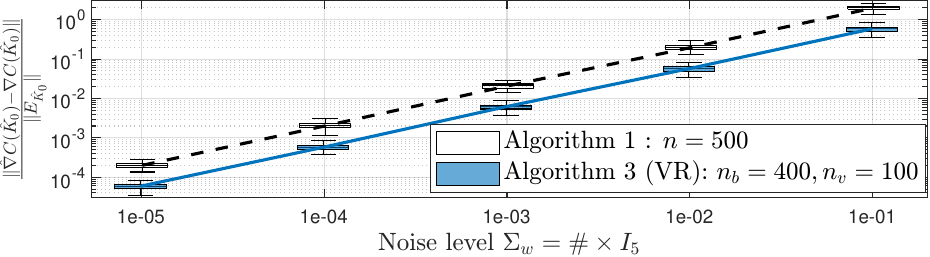}
    \caption{Normalized gradient estimation error as a function of the noise level $\Sigma_w=\#\times I_5$, \textcolor{blue}{for system \eqref{Boeing747} with $\ell=20$ and $r=0.01$}, comparing Algorithm~\ref{Algo1} ($n=500$) and Algorithm~\ref{AlgoVV} with variance reduction ($n_b=400$, $n_v=100$), over 100 Monte Carlo runs shown as box plots.}
    \label{fig:fig5}
\end{figure}
The variance reduction algorithm consistently achieves a smaller estimation error across all noise levels, with the same total sample budget of $n=500$ rollouts, which is consistent with the analysis in Corollary~\ref{Coro1}.
}
\textcolor{black}{
\subsection{Noise Level and Variance Reduction in PGD}
The policy is updated using Algorithm~\ref{Algo2}. Figure~\ref{fig:fig2} illustrates the relative suboptimality gap $\frac{C(\hat{K}_i)-C(K^*)}{C(K^*)}$ as a function of the iteration index $i$, for different noise levels, step sizes, and with/without variance reduction. The parameters $n$, $n_b$, and $n_v$ are set as in the previous subsection (Section \ref{sec:Simulation}\ref{simGE}).
\begin{figure}[H]
    \centering
    \includegraphics[width=1\linewidth]{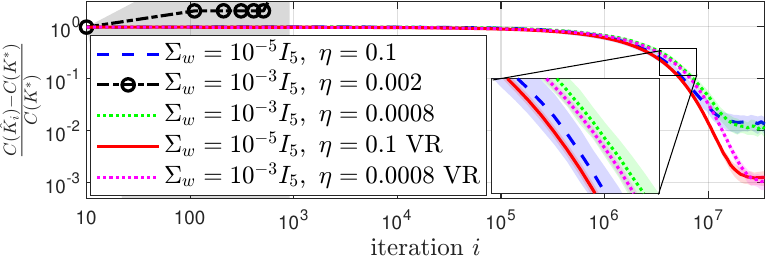}
    \caption{Model-free PGD with and without variance reduction (VR) \textcolor{blue}{for system \eqref{Boeing747} under $\Sigma_w=10^{-5}I$ and $\Sigma_w=10^{-3}I$ with $\ell=20$ and $r=0.01$}, showing mean and standard deviation over 100 Monte Carlo runs. Total rollouts fixed at $n=500$; variance reduction allocates $n_v=100$ and $n_b=400$.}
    \label{fig:fig2}
\end{figure}
For low noise $\Sigma_w=10^{-5}I_5$ with $\eta=0.1$ (blue dashed line), the algorithm converges to a small suboptimality gap arising from the inherent gradient estimation error. When the noise level increases to $\Sigma_w=10^{-3}I_5$, the step size $\eta=0.002$ leads to divergence (black dashed line with circles), consistent with Theorem~\ref{MFPGV}: higher noise requires a smaller step size to ensure convergence. Reducing to $\eta=0.0008$ (green dotted line) recovers convergence at the cost of a slower rate. This behavior is also observed in Figure~\ref{fig:MB}, where decreasing the step size (from the green dash-dot line to the red solid line) leads to convergence. Notably, the application of variance reduction yields a striking improvement: as clearly visible in the figure, the VR variants achieve a substantially smaller optimality gap compared to their non-VR counterparts under the same noise level.
The zoomed inset highlights the benefit of variance reduction. For both noise levels, the VR variants (red solid and magenta dotted lines) converge faster and exhibit significantly smaller variance than their non-VR counterparts, despite using the same total sample budget. This confirms that variance reduction provides a consistent practical improvement.} 

\textcolor{black}{
\subsection{Model-free NPG with Adaptive Step Sizes}\label{Simu2}
The policy is updated using the natural policy gradient method in Algorithm~\ref{Algo3}. Figure~\ref{fig:fig3} shows the relative suboptimality gap for two noise levels and two step size strategies. The fixed step size $\eta_i=\frac{1}{2+25000\,\mathrm{Tr}(P_{\hat{K}_0})}$ is computed once at initialization, while the adaptive step size $\eta_i=\frac{1}{2+25000\,\mathrm{Tr}(P_{\hat{K}_i})}$ is updated at each iteration to reflect the current cost, following the structure of Theorem~\ref{MFNPGV}.
\begin{figure}[h]
 \centering
    \includegraphics[width=1\linewidth]{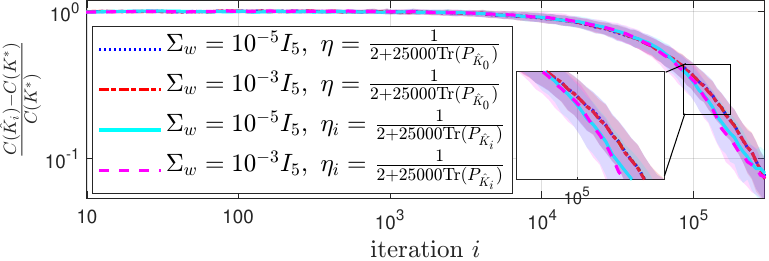}
    \caption{Model-free NPG with fixed and adaptive step sizes \textcolor{blue}{for system \eqref{Boeing747} under $\Sigma_w=10^{-5}I_5$ and $\Sigma_w=10^{-3}I_5$ with $\ell=20$, $r=0.01$, and $n=500$}, showing mean and standard deviation over 100 Monte Carlo runs.}
   \label{fig:fig3}
\end{figure}
\begin{figure}[h]
 \centering
    \includegraphics[width=\linewidth]{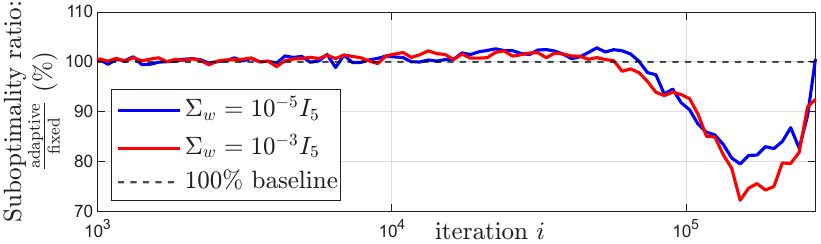}
    \caption{Ratio of the \textcolor{blue}{averaged} suboptimality gap under adaptive step size to that under fixed step size for the model-free NPG (Algorithm~\ref{Algo3}), \textcolor{blue}{for system \eqref{Boeing747} under $\Sigma_w=10^{-5}I_5$ and $\Sigma_w=10^{-3}I_5$ with $\ell=20$, $r=0.01$, and $n=500$}, \textcolor{blue}{i.e. $\mathrm{Ratio}_i=\frac{\overline{C(\hat{K}_i)-C(K^*)}\big|_{\mathrm{adaptive}}}{\overline{C(\hat{K}_i)-C(K^*)}\big|_{\mathrm{fixed}}} (\%)$, where the overline denotes the mean over 100 Monte Carlo runs. A ratio below 100\% indicates that the adaptive step size achieves a smaller suboptimality gap} }
   \label{fig:fig4}
\end{figure}
As shown in the zoomed inset, the adaptive step size variants (cyan solid and magenta dashed lines) converge faster than their fixed step size counterparts (blue dotted and red dash-dotted lines) for both noise levels. To further illustrate the advantage of the adaptive step size, Figure~\ref{fig:fig4} plots the ratio between the suboptimality gap of the adaptive step size and that of the fixed step size for the model-free NPG algorithm. In the early iterations, the ratio remains close to 100\%, since the cost $C(\hat{K}_i)$ has not yet decreased significantly and the adaptive step size $\eta_i$ remains close to $\eta_0$. As the iterations proceed and $C(\hat{K}_i)$ decreases, the adaptive step size grows beyond $\eta_0$, and the ratio drops clearly below 100\% for both noise levels, confirming that the adaptive scheme converges faster in the later stages. Two gaps eventually become comparable again as both algorithms approach the suboptimality floor imposed by the gradient estimation error, at which point the benefit of a larger step size diminishes.\\
Furthermore, for both step size strategies, the two noise levels achieve nearly identical convergence behavior, demonstrating the robustness of the NPG update to varying noise levels.} At higher noise levels, the estimation errors of both $\hat{\nabla}C(\hat{K}_i)$ and $\hat{\Sigma}_{\hat{K}_i}$ grow. These errors are explicitly accounted for in our analysis (Theorems~\ref{Theorem4} and~\ref{Theorem5}). Due to the natural policy gradient update rule (see~\eqref{theorem4.7update}), the estimated gradient is multiplied by the inverse of the estimated covariance, which causes the errors in these two quantities to partially offset each other. If the analysis were performed directly on the product $\hat{\nabla}C(\hat{K}_i)\hat{\Sigma}_{\hat{K}_i}^{-1}$ rather than treating them separately (as in Proof of \textcolor{black}{Theorem \ref{MFNPGV}} in Appendix \ref{ProofMFNPGV}), substantially stronger theoretical guarantees could be established. This observation points to an important direction for future research.
\section{Conclusions}
\label{sec:conclusions}
\textcolor{black}{In this work, we studied the policy gradient methods for the LQR problem with unknown system dynamics subject to stochastic process noise. We established convergence guarantees for both model-based and model-free policy gradient descent and natural policy gradient methods with adaptive step sizes. We further introduced a variance reduction technique with provable sample complexity guarantees, and numerically demonstrated that it improves convergence speed and optimality gap. Future work investigates the model-free Gauss-Newton method, and indirect data-driven policy gradient approaches that integrate online parameter estimation with model-based policy gradient method.}

\section*{APPENDIX}
\section{Proofs}
\subsection{Proof of Theorem \ref{PGTheorem}}\label{Proof1}
  Note that $h_\mathrm{PGD}(C(K))$ is non-increasing in $C(K)$. Following the steps in \cite[Lemma 21]{pmlr-v80-fazel18a}, we can show that if $\eta_i\leq h_\mathrm{PGD}(C(K_i))$, then \textcolor{blue}{$C(K_{i+1})-C(K^*)\leq \gamma_i^p (C(K_{i})-C(K^*))$} for all $i\in \mathbb{Z}_+$, which in turn implies $h_\mathrm{PGD}(C(K_{i+1}))\geq h_\mathrm{PGD}(C(K_i))$. The remainder of the proof follows by setting $\eta_i=\eta_0$ for all $i\in \mathbb{Z}_+$, which yields a lower bound on the convergence rate, and then proceeding as in \cite[Lemma 22]{pmlr-v80-fazel18a}.

\textcolor{blue}{\subsection{Proof of Theorem \ref{NPGTheorem}}\label{Proof2}
  Following the steps in \cite[Lemma 13 and Proof of Theorem 7, Supplementary Material]{pmlr-v80-fazel18a}, we obtain \eqref{PGConvergence}. The remainder of the theorem then follows from the non-increase of the cost $C(K_i)$, together with using the lower bound of step size $\eta_0$.}
\subsection{Proof of Theorem \ref{Theorem4}}\label{ProofTheorem9}
Before starting the proof of the Theorem, we introduce the following Lemmas:
\subsubsection{Approximating $C(K),\Sigma_K$ with finite horizon}\label{Prooffinite}
\begin{Lemma}[Approximating with finite horizon] \label{Lemmafinitness}
For any $K \in \mathcal{S}$, let $\Sigma_K^{(l)}:=\frac{1}{l}\sum_{t=0}^{l-1}\mathbb{E}_{x_0,w_t}\left[x_tx_t^{\top} \right]$ and $C^{(l)}(K):=\frac{1}{l}\sum_{t=0}^{l-1}\mathbb{E}_{x_0,w_t}\left[x_t^{\top}Q_Kx_t \right]$. Given any desired accuracy $\epsilon_{\Sigma},\epsilon_{C}> 0$, if $l\geq l_{\Sigma}$, where
\begin{equation*}\label{l11}
    \textcolor{blue}{l_{\Sigma} :=\frac{C(K)}{\epsilon_{\Sigma}\lambda_1(Q)\lambda_1(\Sigma_w)}\left({\lVert \Sigma_0\rVert}+\frac{C(K)}{\lambda_1(Q)}\right),}
\end{equation*}
then $\lVert \Sigma_K^{(l)}-\Sigma_K \rVert \leq \epsilon_{\Sigma}.$ If $l\geq l_{C}$, where
\begin{equation}\label{ll1}
    \textcolor{blue}{l_C :=\frac{C(K)^2}{\epsilon_{C}\lambda_1^2(\Sigma_w)\lambda_1(Q)}\left({\lVert \Sigma_0\rVert}+\frac{C(K)}{\lambda_1(Q)}\right),}
\end{equation}
then $\lVert C^{(l)}(K)-C(K) \rVert \leq \epsilon_{C}.$
\end{Lemma}
\begin{proof}
\textcolor{blue}{Subtracting the recursion $\mathbb{E}[x_{t+1}x_{t+1}^\top]=A_K\mathbb{E}[x_tx_t^\top]A_K^\top+\Sigma_w$ from the Lyapunov equation $\Sigma_K=A_K\Sigma_KA_K^\top+\Sigma_w$ and applying induction on $t$ gives, for all $t\geq0$,    $\mathbb{E}[x_tx_t^\top]=\Sigma_K+A_K^t(\Sigma_0-\Sigma_K)(A_K^t)^\top,$ so that $\Sigma_K^{(l)}-\Sigma_K=\frac1l\sum_{t=0}^{l-1}A_K^t(\Sigma_0-\Sigma_K)(A_K^t)^\top$. Since $\Sigma_K=A_K\Sigma_KA_K^\top+\Sigma_w\succeq A_K\Sigma_KA_K^\top$, we have $\Sigma_w\preceq\Sigma_K$, and induction on $t$ gives $A_K^t\Sigma_KA_K^{t\top}\preceq\Sigma_K$ and $A_K^t\Sigma_wA_K^{t\top}\preceq\Sigma_K$ for all $t\geq0$; the latter also yields the exact bound $\sum_{t=0}^{l-1}A_K^t\Sigma_wA_K^{t\top}\preceq\sum_{t=0}^\infty A_K^t\Sigma_wA_K^{t\top}=\Sigma_K$. Using $\Sigma_0\preceq\frac{\lVert\Sigma_0\rVert}{\lambda_1(\Sigma_w)}\Sigma_w$ and $\Sigma_K\preceq\frac{\lVert\Sigma_K\rVert}{\lambda_1(\Sigma_w)}\Sigma_w$, and $-\Sigma_K\preceq\Sigma_0-\Sigma_K\preceq\Sigma_0$, we obtain the two-sided bound $-\frac{\lVert\Sigma_K\rVert}{l\lambda_1(\Sigma_w)}\Sigma_K\ \preceq\ \Sigma_K^{(l)}-\Sigma_K\ \preceq\ \frac{\lVert\Sigma_0\rVert}{l\lambda_1(\Sigma_w)}\Sigma_K,$ hence $\lVert \Sigma_K^{(l)}-\Sigma_K\rVert\leq\frac{\lVert\Sigma_K\rVert(\lVert\Sigma_0\rVert+\lVert\Sigma_K\rVert)}{l\lambda_1(\Sigma_w)}$. Since $\lVert\Sigma_K\rVert\leq C(K)/\lambda_1(Q)$, this gives the stated bound on $l_\Sigma$. For the second claim, $|\mathrm{Tr}(Q_K(\Sigma_K^{(l)}-\Sigma_K))|\leq\mathrm{Tr}(Q_K)\lVert\Sigma_K^{(l)}-\Sigma_K\rVert$, and $C(K)=\mathrm{Tr}(Q_K\Sigma_K)\geq\lambda_1(\Sigma_w)\mathrm{Tr}(Q_K)$ (using $\Sigma_K\succeq\Sigma_w$), so $\mathrm{Tr}(Q_K)\leq C(K)/\lambda_1(\Sigma_w)$. Combining with the bound on $\lVert\Sigma_K^{(l)}-\Sigma_K\rVert$ gives the stated bound on $l_C$.}
\end{proof}
\subsubsection{Smoothing and the policy gradient descent analysis}\label{AppSmooth}
The algorithm performs PGD on the following smoothing function \textcolor{blue}{$C_r(K):=\mathbb{E}_{U\sim\mathbb{B}_r}\left[C(K+U)\right]$}, where $\mathbb{B}_r$ is the uniform distribution over all matrices with Frobenius norm less than $r$ (the entire ball):
\begin{Lemma}[Gradient of smoothing function]\textcolor{blue}{\cite[Supplementary Material, Lemma 26]{pmlr-v80-fazel18a}}\label{LemmaSmooth}
\begin{equation*}
    \nabla C_r(K)=\frac{n_xn_u}{r^2}\mathbb{E}_{U\sim\mathbb{S}_r}\left[C(K+U)U\right],
\end{equation*}
where $\mathbb{S}_r$ is the uniform distribution over all matrices with Frobenius norm $r$ (the sphere).
\end{Lemma}

\subsubsection{Estimating $\nabla C(K)$ with finitely many infinite-horizon Rollouts}\label{MC}
In this section, we aim to bound the difference between $\nabla C(K)-\bar{\nabla} C(K)$, where $\bar{\nabla} C(K):=\frac{1}{n}\sum^{n}_{k=1}\frac{n_xn_u}{r^2}C(K+U_k)U_k,$
and $U_k$ is uniformly distributed and $\lVert U_k \rVert_F=r$.
\begin{Lemma}\label{Lemma12}
    Given an accuracy $\epsilon_{nr}$ and probability $\delta_n$, which can be expressed by $\epsilon_r$ and $\epsilon_n$ ($\epsilon_{nr}=\epsilon_r+\epsilon_n$), there exist an exploration radius $r_{\max}(C(K),\epsilon_r)$ and the number of rollouts $N_1(C(K),\epsilon_n,\epsilon_r,\delta_n)$ such that when $r \leq r_{\max}(C(K),\epsilon_r)$ and $n \geq N_1(C(K),\epsilon_n,\epsilon_r,\delta_n)$, then 
    \begin{equation*}
        \mathbb{P}\left\{\lVert \bar{\nabla} C(K)-  \nabla C(K)\rVert \leq \epsilon_{nr} \right\}\geq 1-\delta_n,
    \end{equation*}
    where $r_{\max}:=\min\textcolor{black}{\bigg\{}h(C(K)), \lVert K^* \rVert, \frac{\epsilon_r}{h_{\nabla}(C(K))},\frac{C(K)}{h_CC(K)}\textcolor{black}{\bigg\}}$ with $h,h_{\nabla}$ was defined in Lemma \ref{Lipschitz}; $N_1$ is given in \eqref{Numberofsamples} in Appendix \ref{DetailedExpression}.
 \end{Lemma}
\begin{proof}
\textcolor{black}{
    We decompose the error as $\bar{\nabla} C(K)-\nabla C(K)=(\nabla C_r(K)-\nabla C(K))+(\bar{\nabla} C(K)-\nabla C_r(K))$ and bound each term.
\textbf{First term.} If $r\leq r_{\max}$, Lemma~\ref{Lipschitz} gives $\lVert \nabla C(K+U_k)-{\nabla} C(K) \rVert\leq {\epsilon_r}$ for all admissible $U_k$. Since $\nabla C_r(K)=\mathbb{E}_{U\sim\textcolor{blue}{\mathbb{B}_r}}\left[\nabla C(K+U)\right]$, the triangle inequality gives $\lVert \nabla  C_r(K)-\nabla C(K) \rVert\leq {\epsilon_r}$
\textbf{Second term.} Define $Z'_k:=\frac{n_xn_u}{r^2}C(K+U_k)U_k$ and $Z':=\nabla C_r(K)=\mathbb{E}[Z_k]$. Using $\lVert U_k\rVert\le r$ and $C(K+U_k)\le 2C(K)$, $\lVert Z'_k \rVert\leq \frac{n_xn_u\cdot 2C(K)}{r},~\lVert Z'\rVert\leq \epsilon_r+b_{\nabla}(C(K)).$ Hence $ \lVert Z'_k-Z'\rVert\leq \alpha_2(C(K),\epsilon_r)$ and $\lVert \textcolor{black}{\mathbb{E}[Z_k^{'\top} Z'_k]}-Z'^\top Z'\rVert \leq \alpha_3(C(K),\epsilon_r)$, with $\alpha_2,\alpha_3$ defined in \eqref{alpha5}
By the Matrix Bernstein Inequality \cite[Lemma B5]{9254115}, taking $N_1$ (defined in \eqref{Numberofsamples} in Appendix \ref{DetailedExpression})
rollouts ensures $\mathbb{P}\left[ \lVert \bar{\nabla} C(K)-\nabla C_r(K)  \rVert_F \leq \epsilon_n \right]\geq 1-\delta_n.$
\textbf{Combining both terms:} if $r\leq r_{\max}$ and $n\geq N_1$, then $\mathbb{P}\left[ \lVert \bar{\nabla} C(K)-\nabla C(K)  \rVert_F \leq \epsilon_{nr} \right]\geq 1-\delta_n.$}
\end{proof}
\subsubsection{Boundedness of samples}\label{Proofboundedness}
\textcolor{black}{Define $X:=\frac{1}{\sqrt{l}}[x_0,\dots,x_{l-1}]^\top$, where $x_t$ with $t\in [0,l-1]$ is the state generated by stabilizing gain $K$ at time $t$. From \eqref{covariance}, $X$ is a zero-mean Gaussian vector with covariance
\begin{equation*}
    \Sigma_X:=\frac{1}{l}\begin{bmatrix}
      \Sigma_0   & \cdots & \Sigma_0 A_K^{l-1\top}\\
      \vdots & \ddots & \vdots\\
      A_K^{l-1}\Sigma_0 & \cdots & \Sigma_{l-1}
    \end{bmatrix},
\end{equation*}
where $\lambda_{ln_x}(\Sigma_X)$ denotes the largest eigenvalue of $\Sigma_X\in\mathbb{R}^{ln_x\times ln_x}$. We have $\lVert \Sigma_X\rVert = \lambda_{ln_x}(\Sigma_X)\leq \mathrm{tr}(\Sigma_X)\leq \frac{1}{l}\mathrm{tr}\Big(\sum_{t=0}^{l-1}\Sigma_t\Big)\leq \frac{n_xC(K)}{\lambda_1(Q)}\Big(1+\textcolor{blue}{\frac{\lambda_{n_x}(\Sigma_0)}{\lambda_1(\Sigma_w)}}\Big)=:\bar{\Sigma}_x.$}
Applying the Hanson-Wright inequality \cite{Vershynin_2018} gives:
\begin{Lemma}[Concentration Bounds for Quadratic Forms]\label{boundofstate}
    For any $K\in \mathcal{S}$, $W\succ 0$, and desired probability $\delta_x>0$:
    \begin{equation*}
    \mathbb{P}\Big[\big|X^{\top}WX-\mathbb{E}[X^{\top}WX^{}]\big|\leq \bar{x}\Big]\geq 1-\delta_x,    
    \end{equation*}
    \textcolor{blue}{where $\bar{x}:=\bar{\Sigma}_x\left(\dfrac{\lVert W\rVert_F^2}{4\lVert W\rVert}+16\lVert W\rVert\ln\dfrac{2}{\delta_x}\right)$.}
\end{Lemma}
\textcolor{blue}{The bound follows from the Hanson--Wright inequality together with $\max\{a,b\}\leq a+b$ (for $a,b\geq0$) and the AM--GM inequality. Taking $X$ generated by gain $\bar{K}_k$ and $W=I_l\otimes Q_{\bar K_k}$ turns $X^\top WX=\frac1l\sum_{t=0}^{l-1}x_t^\top Q_{\bar K_k}x_t$ into the empirical cost $\hat C_{\bar K_k}$, whose expectation is the finite-horizon cost $C^{(l)}(\bar K_k)$, so Lemma~\ref{boundofstate} directly bounds their gap. Since the additive noise is Gaussian and hence unbounded, $\hat C_{\bar K_k}$ and $\hat{\Sigma}_k$ from Algorithm \ref{Algo1} can only be bounded with high probability; Lemma~\ref{boundofstate} supplies this bound, needed by Theorems~3 and~4 to invoke the Matrix Bernstein inequality~\cite[Lemma~B5]{9254115}.}
\subsubsection{Proof of \textcolor{black}{Theorem \ref{Theorem4}}}\label{PfMFPGV}
We introduce the intermediate estimate $\nabla' C(K):=\frac{1}{n}\sum^{n}_{k=1}\frac{n_xn_u}{r^2}C^{(l)}(K+U_k)U_k$, where $C^{(l)}(K)$ is defined in Lemma~\ref{Lemmafinitness}, and decompose the total error as: $\hat{\nabla}C(K)-\nabla C(K)=(\hat{\nabla}C(K)-\nabla' C(K))+(\nabla' C(K)-\bar{\nabla} C(K))+(\bar{\nabla} C(K)-\nabla C(K))   $
where $\bar{\nabla}C(K)$ is defined in Lemma~\ref{Lemma12}. We bound each term separately.
\textbf{Third term.} By Lemma~\ref{Lemma12} with $\epsilon_{nr}:=\epsilon_n+\epsilon_r$, there exist $r_{\max}$ as in \eqref{rexpresion} and $N_1$ such that if $r\leq r_{\max}$ and $n\geq N_1$:
$\mathbb{P}\left[\lVert \bar{\nabla} C(K)-\nabla C(K)\rVert_F \leq \epsilon_{nr}\right]\geq 1-\delta_n.$ \textbf{Second term.} By Lemma~\ref{Lemmafinitness}, for any desired $\epsilon_l>0$, choosing $l\geq l_{\min},$ 
\begin{equation}\label{lexpression}
        l_{\min} := \frac{n_xn_u(2C(K))^2}{\epsilon_lr\lambda_1^2(\Sigma_w)\lambda_1(Q)}\left({\lVert \Sigma_0\rVert}+\frac{2C(K)}{\lambda_1(Q)}\right),
\end{equation}
ensures $|C^{(l)}(\bar{K}_k)-C(\bar{K}_k)|\leq \frac{r\epsilon_l}{n_xn_u}$ for all $k\in[1,n]$. By the triangle inequality and $\lVert U_k\rVert_F=r$: $\lVert\nabla' C(K)-\bar{\nabla} C(K)\rVert\leq\epsilon_l.$ \textcolor{blue}{\textbf{First term.} From Lemma~\ref{boundofstate}, for any $\delta_x>0$ and each $k\in[1,n]$, applying the bound
with per-sample confidence $\frac{\delta_x}{n}$:
$\mathbb{P}(\hat{C}_{\bar{K}_k}\leq \bar{C}_k')\geq 1-\frac{\delta_x}{n},~
\bar{C}_k':=\bar{\Sigma}_x\Big(\lVert Q_{\bar{K}_k}\rVert+\dfrac{\lVert Q_{\bar{K}_k}\rVert_F^2}{4\lVert Q_{\bar{K}_k}\rVert}
+16\lVert Q_{\bar{K}_k}\rVert\,\ln\frac{2n}{\delta_x}\Big)$. A union bound over $k=1,\dots,n$ then gives $\bar{C}'_k\leq\bar{C}(n)$ for \emph{all} $k$
simultaneously with probability at least $1-\delta_x$ (with $\bar{C}$ defined in \eqref{VRtest1}), so each sample
$Z_k:=\frac{n_xn_u}{r^2}\hat{C}_{\bar{K}_k}U_k$ satisfies: $\lVert Z_k\rVert\leq\frac{n_xn_u}{r}\bar{C}(n)=:\alpha_4(C(K),\delta_x,n).$}
\textcolor{blue}{Conditioning on $U_1,\dots,U_n$, each $Z_k$ satisfies $\mathbb{E}[Z_k\mid U_k]=\frac{n_xn_u}{r^2}C^{(l)}(\bar{K}_k)U_k$;
setting $Z:=\frac{1}{n}\sum_{k=1}^n\mathbb{E}[Z_k\mid U_k]=\nabla'C(K)$ and using the triangle inequality:}
$\lVert Z\rVert\leq\epsilon_l+\epsilon_{nr}+\lVert\nabla C(K)\rVert\leq\epsilon_l+\epsilon_{nr}+b_{\nabla}(C(K))=:\bar{Z},$ which gives the bounds: \textcolor{blue}{$\lVert Z_k-Z \rVert\leq \bar{Z}+\alpha_4(C(K),\delta_x,n)=:\alpha_5(C(K),\delta_x,n);$ and $ \lVert{\mathbb{E}[Z_kZ_k^\top]}-ZZ^\top\rVert  \leq \max\{n_x,n_u\}^2 {\alpha_4(C(K),\delta_x,n)^2}\quad+\bar{Z}^2=:\alpha_6(C(K),\delta_x,n).$} \textcolor{black}{Applying \cite[Lemma B5]{9254115}} with accuracy $\epsilon_d$ and probability $\delta_d$, taking $n\geq N'_2(n)$ (defined in \eqref{Nexpression2}) ensures:
$\mathbb{P}\big[\lVert\hat{\nabla}C(K)-\nabla'C(K)\rVert_F\leq\epsilon_d\big]\geq 1-\delta_d.$ \textcolor{blue}{Since $N'_2(n)$ depends on $n$ through $\bar C (n)$, the requirement $n\geq N'_2(n)$ is only implicit. However,
$N'_2(n)$ grows only logarithmically in $n$ (through $\ln n$), while the requirement grows linearly in $n$;
hence for $n$ large enough the linear side always overtakes the logarithmic side, and an explicit
threshold $N_2$ can be found directly as in \eqref{Nexpression22}. }
\textbf{Conclusion.} Combining all terms, when $r\leq r_{\max}$, $l\geq l_{\min}$, and $n\geq\max\{N_1,N_2\}$, then $\mathbb{P}[\lVert\hat{\nabla}C(K)-\nabla C(K)\rVert\leq\epsilon]\geq 1-\delta$, with $r_{\max}$:
\begin{equation}\label{rexpresion}
    r_{\max} = \min\textcolor{black}{\bigg\{}h(C(K)), \lVert K^* \rVert, \frac{\epsilon_r}{h_{\nabla}(C(K))},\frac{C(K)}{h_CC(K)}\textcolor{black}{\bigg\}},
\end{equation}
\subsection{Proof of Theorem \ref{Theorem5}}\label{PT5}
The proof follows the same structure as the proof of Theorem~\ref{Theorem4}, with the decomposition: $\hat{\Sigma}_K-\Sigma_K={(\hat{\Sigma}_K-\tilde{\Sigma}^{(l)}_K)}+(\tilde{\Sigma}^{(l)}_K-\tilde{\Sigma}_K)+(\tilde{\Sigma}_K-\Sigma_K),$ where $\tilde{\Sigma}^{(l)}_K:=\frac{1}{n}\sum_{i=1}^n \Sigma^{(l)}_{K+U_i}$ and $\tilde{\Sigma}_K:=\frac{1}{n}\sum_{i=1}^n \Sigma_{K+U_i}$. \textbf{Last term.} By Lemma~\ref{Lipschitz}, choosing $r\leq r'_{\max}$ as 
   \begin{equation}\label{radius2}
        r'_{\max}:=\min\bigg\{h(C(K)), \lVert K^* \rVert , \frac{\epsilon'_r}{h_{\Sigma}(C(K))},\frac{C(K)}{h_CC(K)}\bigg\};
    \end{equation}
 gives $\lVert\tilde{\Sigma}_K-\Sigma_K\rVert\leq\epsilon'_r$. \textbf{Second term.} By Lemma~\ref{Lemmafinitness}, choosing $l\geq l'_{\min}$ as 
 \begin{equation}\label{l2}
    l'_{\min}:=\frac{2C(K)}{\epsilon'_l\lambda_1(\Sigma_w)\lambda_1(Q)}\left({\lVert \Sigma_0\rVert}+\frac{2C(K)}{\lambda_1(Q)}\right)
\end{equation}
 gives $\lVert\tilde{\Sigma}^{(l)}_K-\tilde{\Sigma}_K\rVert\leq\epsilon'_l$. \textcolor{blue}{\textbf{First term.} Defining $\tilde{Z}_k:=\frac{1}{l}\sum_{t=0}^{l-1}x_t^{(k)}x_t^{(k)\top}$, Lemma~\ref{boundofstate}
applied with per-rollout confidence $\delta_x'/n$ gives $\lVert\tilde{Z}_k\rVert\leq\bar{L}(n)$ (defined in \eqref{barL})
with probability $1-\delta_x'/n$ for each $k$; a union bound over $k=1,\dots,n$ gives $\lVert\tilde{Z}_k\rVert\leq\bar{L}(n)$
for all $k$ simultaneously with probability at least $1-\delta_x'$. Conditioning on $U_1,\dots,U_n$, each $\tilde{Z}_k$
satisfies $\mathbb{E}[\tilde{Z}_k\mid U_k]=\Sigma^{(l)}_{K+U_k}$; setting
$\tilde{Z}:=\frac{1}{n}\sum_{k=1}^n\mathbb{E}[\tilde{Z}_k\mid U_k]=\tilde{\Sigma}^{(l)}_K$, we have
$\lVert\tilde{Z}\rVert\leq\frac{2C(K)}{\lambda_1(Q)}$ (which holds deterministically for every realization of $U_k$
with $r\leq r'_{\max}$), yielding: $\lVert\tilde{Z}_k-\tilde{Z}\rVert\leq\alpha_7(C(K),\delta'_x,n)$ and
$\lVert\mathbb{E}[\tilde{Z}_k\tilde{Z}_k^\top]-\tilde{Z}\tilde{Z}^\top\rVert\leq\alpha_8(C(K),\delta'_x,n)$, where $\alpha_7$
and $\alpha_8$ are defined in \eqref{ALPh9}. Applying \cite[Lemma B5]{9254115} with tolerance $\epsilon'_n$ and
probability $\delta'_n$: since $n''_{\min}(n)$ depends on $n$ through $\bar{L}(n)$, the requirement $n\geq n''_{\min}(n)$ is only
implicit; and resolving the resulting dependence of $n''_{\min}$ on $n$ by the same
argument as in the proof of Theorem~\ref{Theorem4}, we obtain $n'_{\min}(n)$ defined in \eqref{n22}, which ensures $\mathbb{P}[\lVert\hat{\Sigma}_K-\tilde{\Sigma}^{(l)}_K\rVert\leq\epsilon'_n]\geq 1-\delta'_n$
and is used in place of the implicit condition in the remainder of the proof.}
\textbf{Conclusion.} A union bound over all three terms gives $\mathbb{P}\{\lVert\hat{\Sigma}_K-\Sigma_K\rVert\leq\epsilon'\}\geq 1-\delta'$.

\subsection{Proof of Theorem \ref{TheoremVariancereduction}}\label{ProofVariance}
    From the proof of Theorem \ref{Theorem4}, the required sample $n_{\min}$ is determined by the upper bound of the sample $\lVert Z_k \rVert$ as $\bar{C}$. Using \eqref{VRtest1}, we know that this upper bound is probabilistically bounded as follows:
    \begin{equation}\label{VR30}
    \begin{split}
               \textcolor{blue}{ \lvert \hat{C}_{\bar{K}_k}-\hat{b}_s\rvert\leq \lvert \hat{C}_{\bar{K}_k}-b_s\rvert+\lvert b_s-\hat{b}_s\rvert.}
    \end{split}
    \end{equation}
    \textcolor{blue}{Using the definition of suboptimal baseline function $b_s$ \eqref{truebaseline} and {Comparing the expressions of $b_s$ and $\bar{C}(n)$ in \eqref{VRtest1}, we have $b_s\leq C(K)\leq\frac{1}{4}\bar{C}(n)$, and consequently $\lvert\hat{C}_{\bar{K}_k}-b_s\rvert\leq\bar{C}-b_s$. If $\lvert b_s-\hat{b}_s\rvert\leq\epsilon_v b_s$ with $\epsilon_v\in (0,1)$, then from \eqref{VR30}: $\lvert\hat{C}_{\bar{K}_k}-\hat{b}_s\rvert\leq\bar{C}(n)-(1-\epsilon_v)b_s,\forall k\in[1,...,n].$}
Since $\epsilon_v\in(0,1)$, the effective upper bound on the sample is strictly smaller than $\bar{C}(n)$, which is the bound used in Theorem~\ref{Theorem4} without variance reduction. Therefore, the required sample for variance reduction is at least smaller than the case without variance reduction. }Following the proof of Theorem \ref{Theorem4}, we can obtain the expression of $N_b$, given in \eqref{n3expression} in Appendix \ref{DetailedExpression}, which is a function of the baseline function estimates. 
\textcolor{blue}{From Algorithm~\ref{AlgoV}, using the same analysis as in Appendix~\ref{ProofTheorem9}.\ref{Proofboundedness}, by the Gaussian quadratic form concentration inequality \cite[Theorem 2.1]{hsu2012tail}, if the number of rollouts satisfies $n_v\geq \tilde{n}_{\min}(C(K),l,{\delta}_v,\epsilon_v)$, where $\tilde{n}_{\min}$ is defined in \eqref{tilden} in Appendix~\ref{DetailedExpression}, then $\mathbb{P}(|\hat{b}_s-b_s|\leq\epsilon_vb_s)\geq 1-{\delta}_v$. Consequently, the sample size required to estimate the gradient with the baseline function is no larger than that without it, with probability at least $1-{\delta}_v$.}
  

\subsection{Proof of Corollary \ref{Coro1}}\label{Proofcoro}
\textcolor{blue}{Define $\Delta N := n_{\min} - N_b(\hat{b}_s)$. From the expressions in Appendix, we obtain $\Delta N\geq 16D_1(\zeta_1)-16D_1(\zeta^V_1)$ where $D_1$ is defined in \eqref{D1}. From \eqref{tilden}, if $\epsilon_vb_s\leq 1$, then $\tilde{n}_{\min} \leq  \bar{N}:= \frac{4\,l\,n_x^3\,C(K)^2(\lambda_{n_x}(Q_K)^2\lambda_1(\Sigma_w)+\lambda_1(Q))}{\lambda_1(Q)^2\lambda_1(\Sigma_w)(\epsilon_vb_s)^2}\Big(1+\frac{\lambda_{n_x}(\Sigma_0)}{\lambda_1(\Sigma_w)}\Big)^2.$ Simplifying $\Delta N$ further yields $\Delta N\geq\frac{32}{\epsilon_d^2}\log\!\left[\frac{2}{\delta_d}\right]\left[2\max\{n_x,n_u\}^2(2\zeta_1-\Delta)\Delta\right]$ with $\Delta:=\zeta_1-\zeta_1^V=\frac{n_xn_u}{r}\big(b_s-\lvert b_s-\hat b_s\rvert\big)$ and $\zeta_1$ defined in \eqref{Nexpression22}. Using $2\zeta_1-\Delta\geq2\zeta_1-\frac{n_xn_u}{r}b_s$ together with $\Delta\geq \frac{n_xn_u (1-\epsilon_v)b_s}{r}$, and the stated conditions on $\epsilon_v$, we conclude $\Delta_N\geq\bar{N}\geq \tilde{n}_{\min}.$}

\textcolor{black}{
\subsection{Proof of Theorem \ref{MFPGV}}\label{ProofMFPG}
Let $K_{i+1}:=\hat{K}_i-\eta_i\nabla C(\hat{K}_i)$ denote the iterate with exact gradient. If $|C(\hat{K}_{i+1})-C(K_{i+1})|\leq\frac{2\eta_i\sigma\lambda_1(R)\lambda_1^2(\Sigma_w)}{\lVert\Sigma_{K^*}\rVert}\epsilon,$ then when $C(\hat{K}_i)-C(K^*)\geq\epsilon$, equation~\eqref{GD} gives: $ C(\hat{K}_{i+1})-C(K^*)\leq \gamma_i(C(\hat{K}_i)-C(K^*)),$ guaranteeing convergence. Since $\hat{K}_{i+1}-K_{i+1}=\eta_i(\hat{\nabla}C(\hat{K}_i)-\nabla C(\hat{K}_i))$, Lipschitz continuity of $C$ gives: $ |C(\hat{K}_{i+1})-C(K_{i+1})|\leq\eta_i h_{\textcolor{blue}{C}}(C(\hat{K}_i))\lVert\hat{\nabla}C(\hat{K}_i)-\nabla C(\hat{K}_i)\rVert.$ The required gradient accuracy is therefore:
 $\lVert\hat{\nabla}C(\hat{K}_i)-\nabla C(\hat{K}_i)\rVert\leq\frac{2\epsilon\sigma\lambda_1(R)\lambda_1^2(\Sigma_w)}{h_{\textcolor{blue}{C}}(C(\hat{K}_i))\lVert\Sigma_{K^*}\rVert}=:\epsilon_\mathrm{PGD}.$
Choosing $r,l,n$ as in Theorem~\ref{Theorem4} to achieve accuracy $\epsilon_\mathrm{PGD}$ and probability $\delta$ at each iteration, the union bound and $(1-\delta)^{n_\mathrm{PGD}}\geq 1-\delta n_\mathrm{PGD}$ give the result. The remainder follows Theorem~\ref{PGTheorem}.}
\textcolor{black}{
\subsection{Proof of Theorem \ref{MFNPGV}}\label{ProofMFNPGV}
Let $K_{i+1}:=\hat{K}_i-\eta_i\nabla C(\hat{K}_i)\Sigma_{\hat{K}_i}^{-1}$ denote the iterate with exact gradient and covariance. By Theorem~\ref{NPGTheorem}, convergence is guaranteed if $\lvert C(\hat{K}_{i+1})-C(K_{i+1})\rvert\leq\frac{2\epsilon\sigma\eta_i\lambda_1(R)\lambda_1(\Sigma_w)}{\lVert\Sigma_{K^*}\rVert}=:\epsilon_{\mathrm{NPG}}$, which gives, when $C(\hat{K}_i)-C(K^*)\geq\epsilon$: $ C(\hat{K}_{i+1})-C(K^*)\leq\kappa_i(C(\hat{K}_i)-C(K^*)).$
We decompose the error using the triangle inequality: $\lVert\hat{\nabla}C(\hat{K}_i)\hat{\Sigma}_{\hat{K}_i}^{-1}-\nabla C(\hat{K}_i)\Sigma_{\hat{K}_i}^{-1}\rVert \leq\lVert\hat{\nabla}C(\hat{K}_i)-\nabla C(\hat{K}_i)\rVert\lVert\hat{\Sigma}_{\hat{K}_i}^{-1}\rVert+\lVert\nabla C(\hat{K}_i)\rVert\lVert\hat{\Sigma}_{\hat{K}_i}^{-1}-\Sigma_{\hat{K}_i}^{-1}\rVert.$
\textbf{First term.} By Weyl's theorem \cite[Theorem 31]{pmlr-v80-fazel18a}, if $\epsilon\leq\frac{\lambda_1(\Sigma_w)}{2}$, then $\lVert\hat{\Sigma}_{\hat{K}_i}^{-1}\rVert\leq\frac{2}{\lambda_1(\Sigma_w)}$. The required gradient accuracy is: $\lVert\hat{\nabla}C(\hat{K}_i)-\nabla C(\hat{K}_i)\rVert\leq\Delta_i^{\nabla}.$
Choose $l^{\nabla}_i,n^{\nabla}_i,r^{\nabla}_i$ in Algorithm~\ref{Algo1} via Theorem~\ref{Theorem4} to achieve accuracy $\Delta_i^\nabla$ with probability ${1-\frac{\delta}{2}}$.
\textbf{Second term.} By matrix perturbation \textcolor{blue}{\cite[Theorem 32]{pmlr-v80-fazel18a}}, if $\lambda_1(\Sigma_{\hat{K}_i})\geq\lambda_1(\Sigma_w)$ and $\lVert\hat{\Sigma}_{\hat{K}_i}-\Sigma_{\hat{K}_i}\rVert\leq\frac{\lambda_1(\Sigma_w)}{2}$, then the required covariance accuracy is: $ \lVert\hat{\Sigma}_{\hat{K}_i}-\Sigma_{\hat{K}_i}\rVert\leq\Delta_{i}^\Sigma.$
Choose \textcolor{blue}{$l^\Sigma_i,n^\Sigma_i,r^\Sigma_i$} via Theorem~\ref{Theorem5} to achieve accuracy $\Delta_{i}^\Sigma$ with probability ${1-\frac{\delta}{2}}$.
\textbf{Conclusion.} Setting $n_i\geq\max\{n^\Sigma_i,n^\nabla_i\}$, $l_i\geq\max\{l^\Sigma_i,l^\nabla_i\}$, $r_i\leq\min\{r^\Sigma_i,r^\nabla_i\}$ and applying a union bound gives: $ \mathbb{P}\left\{\lvert C(\hat{K}_{i+1})-C(K_{i+1})\rvert\leq\epsilon_{\mathrm{NPG}}\right\}\geq 1-\delta.$ The remainder follows Theorem~\ref{NPGTheorem} via the union bound and $(1-\delta)^{n_\mathrm{NPG}}\geq 1-\delta n_\mathrm{NPG}$.}

\begin{figure*}[p]
 \section{Full expressions for quantities introduced throughout the paper}\label{DetailedExpression}  
    \begin{equation}\label{boundedgradienteq}
\begin{split}
         b_{\nabla}(C(K)):=\frac{2C(K)\alpha_1}{\lambda_1(Q)}, b_{K}(C(K)):= \frac{\lVert B \rVert \lVert A \rVert C(K)}{\lambda_1(\Sigma_w)\lambda_1(R) }+ \frac{\alpha_1}{\lambda_1(R) },\alpha_1:=\sqrt{ \frac{(C(K)-C(K^*))}{\textcolor{black}{\lambda_1(\Sigma_w)}}\left(  \lVert R \rVert+ \frac{\lVert B \rVert^2 C(K)}{{\lambda_1{(\Sigma_w)}}}\right)}
\end{split}
\end{equation}
\textcolor{blue}{
  \begin{equation}\label{Pertubation1}
    \begin{split}
      h_{C}(C(K)) := 2\left( \frac{C(K)}{\lambda_1(\Sigma_w)\lambda_1(Q)}\right)^2\lVert R \rVert\left( b^2_{K}(C(K))\lVert B \rVert(\lVert A\rVert+\lVert B \rVert b_{K}(C(K))+1)+2b_{K}(C(K))+\lVert K^* \rVert\right)\mathrm{Tr}(\Sigma_w),
    \end{split}
\end{equation}
\begin{equation}\label{ErrorGradient}
    h_{\nabla}(C(K)):= 2h_{\Sigma}{(C(K))}\alpha_1+2\lVert R \rVert+ \frac{\lVert B \rVert^2 C(K)}{\lambda_1{(\Sigma_w)}}+\frac{h_C(C(K))}{\mathrm{Tr}(\Sigma_w)}\left(  \lVert B \rVert\lVert A \rVert+(b_{K}(C(K))+\lVert K^*\rVert) \lVert B \rVert^2 \right),
\end{equation}}
  \begin{equation}\label{pG2}
        h_{\mathrm{PGD}}(C(K)):={\frac{1}{32}\min \bigg\{ \left( \frac{\lambda_1(Q)\lambda_1(\Sigma_w)}{C(K)} \right)^2\frac{1}{\lVert B \rVert b_{\nabla}(C(K))(\lVert A\rVert+\lVert B\rVert b_K(C(K)))}, \frac{\lambda_1(Q)}{2C(K)\left(  \lVert R \rVert+ \frac{\lVert B \rVert^2 C(K)}{\lambda_1{(\Sigma_w)}}\right)}\bigg\}},
    \end{equation}

\begin{equation}\label{Numberofsamples}
    N_1:=\frac{2\min\{n_x,n_u\}}{\epsilon_n^2} \left( \alpha_3(C(K),\epsilon_r) +\frac{\alpha_2(C(K),\epsilon_r)\epsilon_n}{3\sqrt{\min(n_x,n_u)}}\right)\log \left[ \frac{n_x+n_u}{\delta_n}\right],
\end{equation}
\begin{equation}\label{alpha5}
    \begin{split}
  \alpha_2(C(K),\epsilon_r):=\frac{2n_xn_uC(K)}{r} +\epsilon_r+b_{\nabla}(C(K)),~\alpha_3(C(K),\epsilon_r):= \left(\frac{2\max\{n_x,n_u\}n_xn_uC(K)}{r}\right)^2+ \left( \epsilon_r+b_{\nabla}(C(K))\right)^2
    \end{split}
\end{equation}

\begin{equation}\label{VRtest1}
    \textcolor{blue}{\bar{C}(n):=\zeta \left(1+\frac{n_x}{4}+16\ln\frac{2n}{\delta_x}\right)~\mathrm{with~} \zeta:=\frac{(2C(K))^2ln_x}{\lambda_1(Q)\lambda_1(\Sigma_w)}\left(1+\frac{\lambda_{n_x}(\Sigma_0)}{\lambda_1(\Sigma_w)}\right)}
\end{equation}
\textcolor{blue}{
\begin{equation}\label{Nexpression2}
N'_2(\textcolor{blue}{n}):=\frac{2\min\{n_x,n_u\}}{\epsilon_d^2} \left( \alpha_6(C(K),\delta_x,\textcolor{blue}{n}) +\frac{\alpha_5(C(K),\delta_x,\textcolor{blue}{n})\epsilon_d}{3\sqrt{\min(n_x,n_u)}}\right)\log \left[ \frac{n_x+n_u}{\delta_d}\right],
\end{equation}
\begin{equation}\label{Nexpression22}
    N_2:=16\,D_1(\zeta_1)\big(D+\ln(16D_1(\zeta_1))\big)^2,~\mathrm{with~} D:=1+\ln\frac{2}{\delta_x} ~\mathrm{and~}\zeta_r:=\frac{n_xn_u}{r}\zeta, ~\zeta_1:=\zeta_r\left(1+\frac{n_x}{4}\right)
\end{equation}
\begin{equation}\label{D1}
    D_1(\zeta_1):=\frac{2\min\{n_x,n_u\}}{\epsilon_d^2}\log\!\left[\frac{n_x+n_u}{\delta_d}\right]
\left[2\max\{n_x,n_u\}^2\Big[\zeta_1^2+256\zeta_r^2\Big]+\bar{Z}^2
+\frac{\epsilon_d}{3\sqrt{\min(n_x,n_u)}}\Big(\bar{Z}\zeta_1+16\zeta_r\Big)\right]
\end{equation}
\begin{equation}\label{barL}
\textcolor{blue}{\bar{L}(n):=\zeta_L\left(1+\frac{n_x}{4}+16\ln\frac{2n}{\delta_x}\right)~\mathrm{with~}
\zeta_L:=\frac{2ln_xC(K)}{\lambda_1(Q)}\left(1+\frac{\lambda_{n_x}(\Sigma_0)}{\lambda_1(\Sigma_w)}\right),}
\end{equation}
\begin{equation}\label{ALPh9}
    \textcolor{blue}{\alpha_7(C(K),\delta'_x,n):=\frac{2C(K)}{\lambda_1(Q)}+\bar{L}(n)~\mathrm{and}~\alpha_8(C(K),\delta'_x,n):={n_x^2\left[ \bar{L}(n)^2+\left(\frac{2C(K)}{\lambda_1(Q)} \right)^2 \right]},}
\end{equation}
\begin{equation}\label{n2}
    \textcolor{blue}{n_{\min}''(n):= \frac{2n_x}{(\epsilon'_n)^2} \left( \alpha_8(C(K),\delta'_x.n)+\frac{\alpha_7(C(K),\delta'_x,n)\epsilon'_n}{3 \sqrt{n_x}}\right) \log \left[ \frac{2n_x}{\delta'_n}\right],}
\end{equation}
\begin{equation}\label{n22}
    n'_{\min}:=16\,D_1'\big(D'+\ln(16D_1')\big)^2, \mathrm{with}~D':=1+\ln\frac{2}{\delta_x'}
\end{equation}
\begin{equation*}
    D_1':=\frac{2n_x}{(\epsilon_n')^2}\log\!\left[\frac{2n_x}{\delta_n'}\right]
\left[n_x^2\left(2\zeta_L^2\Big(1+\tfrac{n_x}{4}\Big)^2+\Big(\frac{2C(K)}{\lambda_1(Q)}\Big)^2\right)+512\,n_x^2\zeta_L^2
+\frac{\epsilon_n'}{3\sqrt{n_x}}\left(\frac{2C(K)}{\lambda_1(Q)}+\zeta_L\Big(1+\tfrac{n_x}{4}\Big)+16\zeta_L\right)\right]
\end{equation*}
\begin{equation}\label{n3expression}
    N_b:=16\,D_1(\zeta_1^V)\big(D+\ln(16D_1(\zeta_1^V))\big)^2,~\mathrm{with~} \zeta^V_1:=\zeta_r\left(1+\frac{n_x}{4}\right)-\frac{n_xn_u}{r}(b_s-\lvert b_s-\hat{b}_s\rvert)~\mathrm{and~} D_1 \mathrm{~defined~in~\eqref{D1}}
\end{equation}
\begin{equation}\label{tilden}
   \tilde{n}_{\min} := \left(\frac{4\,l\,n_x^3\,C(K)^2\lVert Q+K^\top RK \rVert_F^2}{\lambda_1(Q)^2(\epsilon_vb_s)^2}\Big(1+\frac{\lambda_{n_x}(\Sigma_0)}{\lambda_1(\Sigma_w)}\Big)^2
+\frac{4\,n_xC(K)^2}{\lambda_1(Q)\lambda_1(\Sigma_w)\epsilon_vb_s}\Big(1+\frac{\lambda_{n_x}(\Sigma_0)}{\lambda_1(\Sigma_w)}\Big)\right)\log\frac{2}{\delta_v}
\end{equation}}
\end{figure*}

\section*{ACKNOWLEDGMENT}
We would like to thank Nicolas Chatzikiriakos for his constructive discussion and valuable feedback.
\bibliographystyle{unsrt}  
\bibliography{ref}  

\begin{thebibliography}{10}

\bibitem{Annaswamy_23_ARCRAS}
Anuradha~M. Annaswamy.
\newblock Adaptive control and intersections with reinforcement learning.
\newblock {\em Annual Review of Control, Robotics, and Autonomous Systems}, 6(1):65--93, 2023.

\bibitem{bertsekas2019reinforcement}
D.~Bertsekas.
\newblock {\em Reinforcement Learning and Optimal Control}.
\newblock Athena Scientific optimization and computation series. Athena Scientific, 2019.

\bibitem{Sutton1998}
Richard~S. Sutton and Andrew~G. Barto.
\newblock {\em Reinforcement Learning: An Introduction}.
\newblock The MIT Press, second edition, 2018.

\bibitem{annurev:/content/journals/10.1146/annurev-control-042920-020021}
Bin Hu, Kaiqing Zhang, Na~Li, Mehran Mesbahi, Maryam Fazel, and Tamer Başar.
\newblock Toward a theoretical foundation of policy optimization for learning control policies.
\newblock {\em Annual Review of Control, Robotics, and Autonomous Systems}, 6(Volume 6, 2023):123--158, 2023.

\bibitem{10.5555/3009657.3009806}
Richard~S. Sutton, David McAllester, Satinder Singh, and Yishay Mansour.
\newblock Policy gradient methods for reinforcement learning with function approximation.
\newblock In {\em Proceedings of the 12th International Conference on Neural Information Processing Systems}, page 1057–1063. MIT Press, 1999.

\bibitem{NIPS1999_6449f44a}
Vijay Konda and John Tsitsiklis.
\newblock Actor-critic algorithms.
\newblock In {\em Advances in Neural Information Processing Systems}, pages 1008--13. MIT Press, 1999.

\bibitem{schulman2017proximalpolicyoptimizationalgorithms}
John Schulman, Filip Wolski, Prafulla Dhariwal, Alec Radford, and Oleg Klimov.
\newblock Proximal policy optimization algorithms.
\newblock arXiv preprint arXiv:1707.06347, 2017.

\bibitem{cen2023globalconvergencepolicygradient}
Shicong Cen and Yuejie Chi.
\newblock Global convergence of policy gradient methods in reinforcement learning, games and control.
\newblock arXiv preprint arXiv:2310.05230, 2023.

\bibitem{pmlr-v80-fazel18a}
Maryam Fazel, Rong Ge, Sham Kakade, and Mehran Mesbahi.
\newblock Global convergence of policy gradient methods for the linear quadratic regulator.
\newblock In {\em Proceedings of the 35th International Conference on Machine Learning}, pages 1467--1476. PMLR, 2018.

\bibitem{9254115}
Benjamin Gravell, Peyman~Mohajerin Esfahani, and Tyler Summers.
\newblock Learning optimal controllers for linear systems with multiplicative noise via policy gradient.
\newblock {\em IEEE Transactions on Automatic Control}, 66(11):5283--5298, 2021.

\bibitem{doi:10.1137/20M1347942}
Kaiqing Zhang, Bin Hu, and Tamer Ba\c{s}ar.
\newblock Policy optimization for $\mathcal{H}\_2$ linear control with $\mathcal{H}\_\infty$ robustness guarantee: Implicit regularization and global convergence.
\newblock {\em SIAM Journal on Control and Optimization}, 59(6):4081--4109, 2021.

\bibitem{doi:10.1137/19M1288012}
Kaiqing Zhang, Alec Koppel, Hao Zhu, and Tamer Ba\c{s}ar.
\newblock Global convergence of policy gradient methods to (almost) locally optimal policies.
\newblock {\em SIAM Journal on Control and Optimization}, 58(6):3586--3612, 2020.

\bibitem{10005813}
Feiran Zhao, Keyou You, and Tamer Başar.
\newblock Global convergence of policy gradient primal–dual methods for risk-constrained {LQR}s.
\newblock {\em IEEE Transactions on Automatic Control}, 68(5):2934--2949, 2023.

\bibitem{9992612}
Ingvar Ziemann, Anastasios Tsiamis, Henrik Sandberg, and Nikolai Matni.
\newblock How are policy gradient methods affected by the limits of control?
\newblock In {\em 2022 IEEE 61st Conference on Decision and Control (CDC)}, pages 5992--5999, 2022.

\bibitem{doi:10.1137/23M1560781}
Yinbin Han, Meisam Razaviyayn, and Renyuan Xu.
\newblock Policy gradient converges to the globally optimal policy for nearly linear-quadratic regulators.
\newblock {\em SIAM Journal on Control and Optimization}, 63(4):2936--2963, 2025.

\bibitem{pmlr-v54-abeille17b}
Marc Abeille and Alessandro Lazaric.
\newblock {Thompson Sampling for Linear-Quadratic Control Problems}.
\newblock In {\em Proceedings of the 20th International Conference on Artificial Intelligence and Statistics}, pages 1246--1254. PMLR, 2017.

\bibitem{articlesimulation}
Sarah Dean, Horia Mania, Nikolai Matni, Benjamin Recht, and Stephen Tu.
\newblock On the sample complexity of the linear quadratic regulator.
\newblock {\em Foundations of Computational Mathematics}, 20, 10 2017.

\bibitem{JMLR:v24:21-0842}
Ben Hambly, Renyuan Xu, and Huining Yang.
\newblock Policy gradient methods find the nash equilibrium in n-player general-sum linear-quadratic games.
\newblock {\em Journal of Machine Learning Research}, 24(139):1--56, 2023.

\bibitem{8558117}
Donghwan Lee and Jianghai Hu.
\newblock Primal-dual {Q}-learning framework for {LQR} design.
\newblock {\em IEEE Transactions on Automatic Control}, 64(9):3756--3763, 2019.

\bibitem{10384256}
Victor~G. Lopez and Matthias~A. Müller.
\newblock An efficient off-policy reinforcement learning algorithm for the continuous-time {LQR} problem.
\newblock In {\em 2023 62nd IEEE Conference on Decision and Control (CDC)}, pages 13--19, 2023.

\bibitem{annurev:/content/journals/10.1146/annurev-control-053018-023825}
Benjamin Recht.
\newblock A tour of reinforcement learning: The view from continuous control.
\newblock {\em Annual Review of Control, Robotics, and Autonomous Systems}, 2(Volume 2, 2019):253--279, 2019.

\bibitem{pmlr-v80-tu18a}
Stephen Tu and Benjamin Recht.
\newblock Least-squares temporal difference learning for the linear quadratic regulator.
\newblock In {\em Proceedings of the 35th International Conference on Machine Learning}, pages 5005--5014. PMLR, 2018.

\bibitem{pmlr-v19-abbasi-yadkori11a}
Yasin Abbasi-Yadkori and Csaba Szepesv\'ari.
\newblock Regret bounds for the adaptive control of linear quadratic systems.
\newblock In {\em Proceedings of the 24th Annual Conference on Learning Theory}, pages 1--26. PMLR, 2011.

\bibitem{pmlr-v119-cassel20a}
Asaf Cassel, Alon Cohen, and Tomer Koren.
\newblock Logarithmic regret for learning linear quadratic regulators efficiently.
\newblock In {\em Proceedings of the 37th International Conference on Machine Learning}, pages 1328--1337. PMLR, 2020.

\bibitem{pmlr-v119-simchowitz20a}
Max Simchowitz and Dylan Foster.
\newblock Naive exploration is optimal for online {LQR}.
\newblock In {\em Proceedings of the 37th International Conference on Machine Learning}, pages 8937--8948. PMLR, 2020.

\bibitem{https://doi.org/10.1002/rnc.7475}
Bowen Song and Andrea Iannelli.
\newblock The role of identification in data-driven policy iteration: A system theoretic study.
\newblock {\em International Journal of Robust and Nonlinear Control}, 2024.

\bibitem{10383604}
Lorenzo Sforni, Guido Carnevale, Ivano Notarnicola, and Giuseppe Notarstefano.
\newblock On-policy data-driven linear quadratic regulator via combined policy iteration and recursive least squares.
\newblock In {\em 2023 62nd IEEE Conference on Decision and Control (CDC)}, pages 5047--5052, 2023.

\bibitem{SongIannelli+2025+398+412}
Bowen Song and Andrea Iannelli.
\newblock Robustness of online identification-based policy iteration to noisy data.
\newblock {\em at - Automatisierungstechnik}, 73(6):398--412, 2025.

\bibitem{zhao2025policygradientadaptivecontrol}
Feiran Zhao, Alessandro Chiuso, and Florian Dörfler.
\newblock Policy gradient adaptive control for the {LQR}: Indirect and direct approaches.
\newblock arXiv preprint arXiv:2505.03706, 2025.

\bibitem{11030798}
Feiran Zhao, Alessandro Chiuso, and Florian Dörfler.
\newblock Regularization for covariance parameterization of direct data-driven {LQR} control.
\newblock {\em IEEE Control Systems Letters}, 9:961--966, 2025.

\bibitem{carnevale2025datadrivenlqrfinitetimeexperiments}
Guido Carnevale, Nicola Mimmo, and Giuseppe Notarstefano.
\newblock Data-driven {LQR} with finite-time experiments via extremum-seeking policy iteration.
\newblock arXiv preprint arXiv:2412.02758, 2025.

\bibitem{9448427}
Hesameddin Mohammadi, Armin Zare, Mahdi Soltanolkotabi, and Mihailo~R. Jovanović.
\newblock Convergence and sample complexity of gradient methods for the model-free linear–quadratic regulator problem.
\newblock {\em IEEE Transactions on Automatic Control}, 67(5):2435--2450, 2022.

\bibitem{10091214}
Shokichi Takakura and Kazuhiro Sato.
\newblock Structured output feedback control for linear quadratic regulator using policy gradient method.
\newblock {\em IEEE Transactions on Automatic Control}, 69(1):363--370, 2024.

\bibitem{10669082}
Feiran Zhao, Xingyun Fu, and Keyou You.
\newblock Convergence and sample complexity of policy gradient methods for stabilizing linear systems.
\newblock {\em IEEE Transactions on Automatic Control}, 70(3):1455--1466, 2025.

\bibitem{pmlr-v89-malik19a}
Dhruv Malik, Ashwin Pananjady, Kush Bhatia, Koulik Khamaru, Peter Bartlett, and Martin Wainwright.
\newblock Derivative-free methods for policy optimization: Guarantees for linear quadratic systems.
\newblock In {\em Proceedings of the Twenty-Second International Conference on Artificial Intelligence and Statistics}, volume~89, pages 2916--2925. PMLR, 2019.

\bibitem{li2025robustnessderivativefreemethodslinear}
Weijian Li, Panagiotis Kounatidis, Zhong-Ping Jiang, and Andreas~A. Malikopoulos.
\newblock On the robustness of derivative-free methods for linear quadratic regulator.
\newblock arXiv preprint arXiv:2506.12596, 2025.

\bibitem{moghaddam2025samplecomplexitylinearquadratic}
Amirreza~Neshaei Moghaddam, Alex Olshevsky, and Bahman Gharesifard.
\newblock Sample complexity of the linear quadratic regulator: A reinforcement learning lens.
\newblock arXiv preprint arXiv:2404.10851, 2025.

\bibitem{doi:10.1137/20M1382386}
Ben Hambly, Renyuan Xu, and Huining Yang.
\newblock Policy gradient methods for the noisy linear quadratic regulator over a finite horizon.
\newblock {\em SIAM Journal on Control and Optimization}, 59(5):3359--3391, 2021.

\bibitem{doi:10.1137/23M1554771}
Caleb Ju, Georgios Kotsalis, and Guanghui Lan.
\newblock A model-free first-order method for linear quadratic regulator with \(\boldsymbol{\tilde {o}(1/\varepsilon )}\) sampling complexity.
\newblock {\em SIAM Journal on Control and Optimization}, 63(3):2098--2123, 2025.

\bibitem{NEURIPS2019_9713faa2}
Zhuoran Yang, Yongxin Chen, Mingyi Hong, and Zhaoran Wang.
\newblock Provably global convergence of actor-critic: A case for linear quadratic regulator with ergodic cost.
\newblock In {\em Advances in Neural Information Processing Systems}, volume~32. Curran Associates, Inc., 2019.

\bibitem{pmlr-v99-tu19a}
Stephen Tu and Benjamin Recht.
\newblock The gap between model-based and model-free methods on the linear quadratic regulator: An asymptotic viewpoint.
\newblock In {\em Proceedings of the Thirty-Second Conference on Learning Theory}, pages 3036--3083. PMLR, 2019.

\bibitem{9691800}
Farnaz~Adib Yaghmaie, Fredrik Gustafsson, and Lennart Ljung.
\newblock Linear quadratic control using model-free reinforcement learning.
\newblock {\em IEEE Transactions on Automatic Control}, 68(2):737--752, Feb 2023.

\bibitem{lewis2012optimal}
F.L. Lewis, D.~Vrabie, and V.L. Syrmos.
\newblock {\em Optimal Control}.
\newblock EngineeringPro collection. Wiley, 2012.

\bibitem{NIPS2001_4b86abe4}
Sham~M Kakade.
\newblock A natural policy gradient.
\newblock In {\em Advances in Neural Information Processing Systems}. MIT Press, 2001.

\bibitem{doi:10.1137/1.9780898718768}
Andrew~R. Conn, Katya Scheinberg, and Luis~N. Vicente.
\newblock {\em Introduction to Derivative-Free Optimization}.
\newblock Society for Industrial and Applied Mathematics, 2009.

\bibitem{9186148}
Sijia Liu, Pin-Yu Chen, Bhavya Kailkhura, Gaoyuan Zhang, Alfred~O. Hero~III, and Pramod~K. Varshney.
\newblock A primer on zeroth-order optimization in signal processing and machine learning: Principals, recent advances, and applications.
\newblock {\em IEEE Signal Processing Magazine}, 37(5):43--54, 2020.

\bibitem{6392457}
Ivo Grondman, Lucian Busoniu, Gabriel A.~D. Lopes, and Robert Babuska.
\newblock A survey of actor-critic reinforcement learning: Standard and natural policy gradients.
\newblock {\em IEEE Transactions on Systems, Man, and Cybernetics}, 42(6):1291--1307, 2012.

\bibitem{6315022}
Thomas Degris, Patrick~M. Pilarski, and Richard~S. Sutton.
\newblock Model-free reinforcement learning with continuous action in practice.
\newblock In {\em 2012 American Control Conference (ACC)}, pages 2177--2182, 2012.

\bibitem{pmlr-v48-mniha16}
Volodymyr Mnih, Adria~Puigdomenech Badia, Mehdi Mirza, Alex Graves, Timothy Lillicrap, Tim Harley, David Silver, and Koray Kavukcuoglu.
\newblock Asynchronous methods for deep reinforcement learning.
\newblock In {\em Proceedings of The 33rd International Conference on Machine Learning}, pages 1928--1937. PMLR, 2016.

\bibitem{Vershynin_2018}
Roman Vershynin.
\newblock {\em High-Dimensional Probability: An Introduction with Applications in Data Science}.
\newblock Cambridge Series in Statistical and Probabilistic Mathematics. Cambridge University Press, 2018.

\bibitem{hsu2012tail}
Daniel Hsu, Sham~M. Kakade, and Tong Zhang.
\newblock A tail inequality for quadratic forms of subgaussian random vectors.
\newblock {\em Electronic Communications in Probability}, 17:1--6, 2012.

\end{thebibliography}

\begin{IEEEbiography}[{\includegraphics[width=1in,height=1.25in,clip,keepaspectratio]{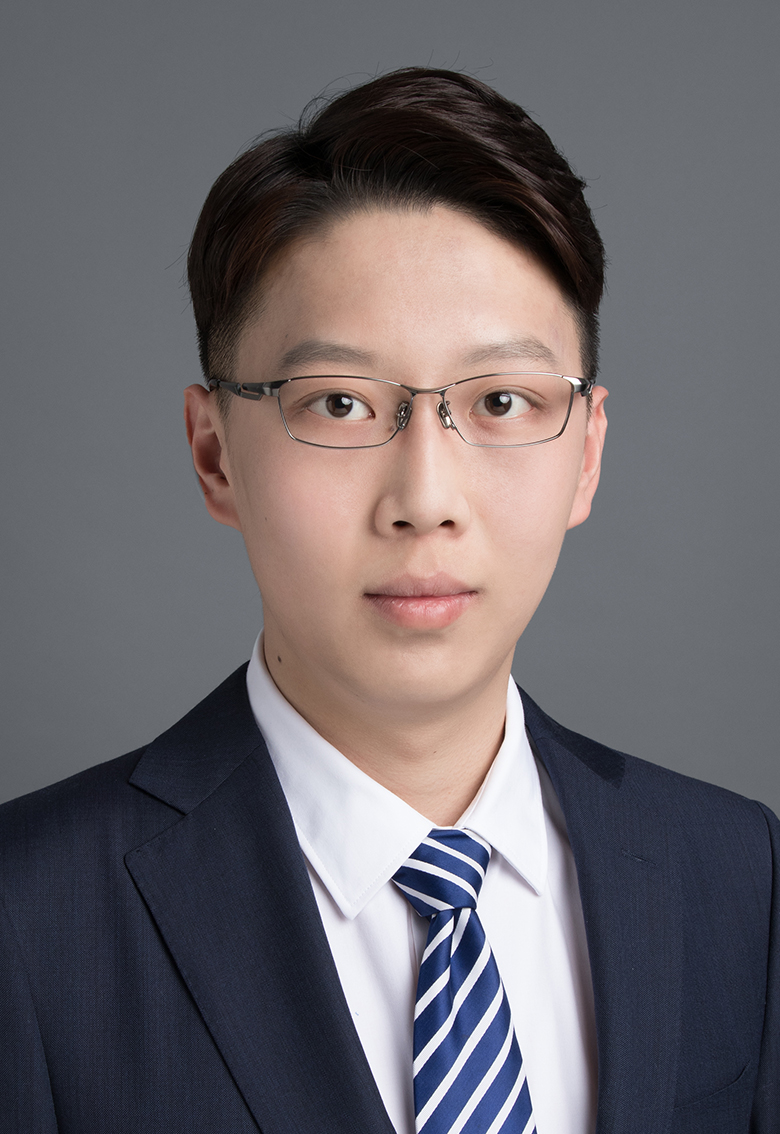}}]{Bowen Song}{\space} is a Ph.D. student at the Institute for Systems Theory and Automatic Control, University of Stuttgart (Germany). He received his B.Eng. in Mechatronics from Tongji University (Shanghai, China) and his M.Sc. in Electrical Engineering and Information Technology from the Technical University of Munich (Germany). He is currently pursuing his Ph.D. in Control Theory and Learning. His research interests include Policy Gradient Methods, Data-driven Control and Reinforcement Learning.
\end{IEEEbiography}

\begin{IEEEbiography}[{\includegraphics[width=1in,height=1.25in,clip,keepaspectratio]{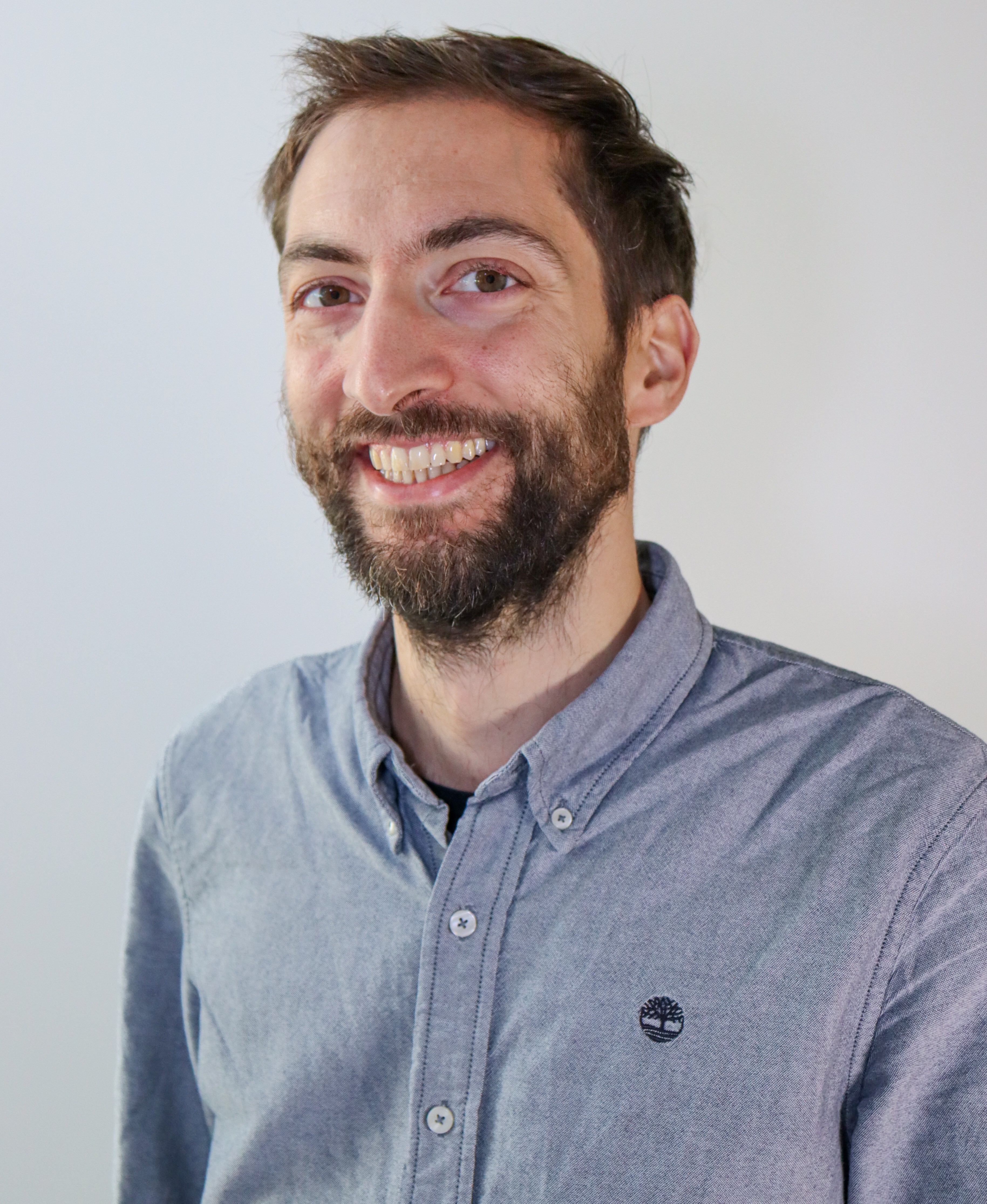}}]{Andrea Iannelli}{\space}(Member, IEEE) is an Assistant Professor in the Institute for Systems Theory and Automatic Control at the University of Stuttgart (Germany). He completed his B.Sc. and M.Sc. degrees in Aerospace Engineering at the University of Pisa (Italy) and received his PhD from the University of Bristol (United Kingdom) on robust control and dynamical systems theory. He was a postdoctoral researcher in the Automatic Control Laboratory at ETH Zurich (Switzerland). His main research interests are at the intersection of control theory, optimization, and learning, with a particular focus on robust and adaptive optimization-based control, uncertainty quantification, and sequential decision-making problems. He serves the community as Associated Editor for the International Journal of Robust and Nonlinear Control and as IPC member of international conferences in the areas of control, optimization, and learning.
\end{IEEEbiography}

\end{document}